\documentclass[12pt]{article}
\usepackage{graphicx}
\usepackage{amssymb,textcomp,amsmath}
\long\def\symbolfootnote[#1]#2{\begingroup
\def\thefootnote{\fnsymbol{footnote}}
\footnote[#1]{#2}\endgroup}
\def\lsim{\:\raisebox{-0.75ex}{$\stackrel{\textstyle<}{\sim}$}\:}
\def\gsim{\:\raisebox{-0.75ex}{$\stackrel{\textstyle>}{\sim}$}\:}

\numberwithin{equation}{section}

\begin{document}
 
\begin{center}
 \Large{\bf Running Spectral Index and Formation of Primordial Black Hole
   in Single Field Inflation Models}
\end{center}
\begin{center}
\large{Manuel Drees$^*$} and \large{Encieh Erfani$^\dag$}
\end{center}
\begin{center}
 \textit{Physikalisches Institut and Bethe Center for Theoretical Physics, 
 Universit\"{a}t Bonn,\\ Nussallee 12, 53115 Bonn, Germany}
\end{center}

\date{}

\symbolfootnote[0]{$^{*}$drees@th.physik.uni-bonn.de}
\symbolfootnote[0]{$^{\dag}$erfani@th.physik.uni-bonn.de}

\begin{abstract}
  A broad range of single field models of inflation are analyzed in
  light of all relevant recent cosmological data, checking whether
  they can lead to the formation of long--lived Primordial Black Holes
  (PBHs). To that end we calculate the spectral index of the power
  spectrum of primordial perturbations as well as its first and second
  derivatives. PBH formation is possible only if the spectral index
  increases significantly at small scales, i.e. large wave number
  $k$. Since current data indicate that the first derivative
  $\alpha_S$ of the spectral index $n_S(k_0)$ is negative at the
  pivot scale $k_0$, PBH formation is only possible in the presence of
  a sizable and positive second derivative (``running of the
  running'') $\beta_S$. Among the three small--field and five
  large--field models we analyze, only one small--field model, the
  ``running mass'' model, allows PBH formation, for a narrow range of
  parameters. We also note that none of the models we analyze can
  accord for a large and negative value of $\alpha_S$, which is weakly
  preferred by current data.
\end{abstract}
\newpage

\section{Introduction}

An epoch of accelerated expansion in the early universe, inflation,
dynamically resolves many cosmological puzzles of the hot big bang
model, such as homogeneity, isotropy and flatness of the Universe
\cite{kt}. On the other hand, the generation of a spectrum of
primordial fluctuations in the early universe is a crucial ingredient
of all inflationary models. These fluctuations can explain the
generation of all (classical) inhomogeneities that can be seen in our
universe, from the Cosmic Microwave Background (CMB) anisotropies to
the Large Scale Structure (LSS) in the form of (clusters of) galaxies
\cite{Lyth Book}. The inflationary paradigm therefore reconciles Big
Bang cosmology with the appearance of an inhomogeneous universe. In
addition, many models of inflation \cite{review} make accurate
predictions allowing for observations of ever increasing variety and
quality to discriminate between the various candidates.

One such prediction concerns the possible formation of Primordial
Black Holes (PBHs). Indeed, it was realized already some time ago that
the spectrum of primordial fluctuations could lead to the production
of PBHs \cite{Carr1,Carr2}. For this generation mechanism to be
efficient, one typically needs a ``blue'' spectrum
\cite{PBH_formation,encieh}. This means that the amplitude of density
fluctuations must be much higher at the small length scales relevant
for PBH formation than at the much larger length scales probed by data
on the Cosmic Microwave Background (CMB) and Large Scale Structure
(LSS). This discrepancy of scales makes PBH formation also a unique
probe of cosmological inflation; in particular, the constraint that
not too many PBHs should have been produced has been used to limit the
power spectrum at small length scales, which in turn allows to
constrain models of inflation \cite{pbh_constraint,earlyPBH}. For
single--field inflation models, the relevant parameter space is
defined by the scalar spectral index $n_S$; the ratio of tensor to
scalar fluctuations $r$; the running of the scalar spectral index
$\alpha_S$; and we will see that the ``running of running of the
spectral index'', $\beta_S$, is also important for PBH formation.

The goal of this paper is to make use of the recent observational
bounds on these parameters derived from the combined CMB data from the
Wilkinson Microwave Anisotropy Probe seven year data (WMAP7)
\cite{WMAP7} and the South Pole Telescope (SPT) \cite{SPT}, as well as
data on the Baryon Acoustic Oscillations (BAO) \cite{BAO}, the Hubble
constant $H_0$ \cite{H0}, and clusters of galaxies \cite{CLUSTERS}. We
will investigate a wide range of inflationary models, checking whether
they still can give rise to significant PBH formation given these
constraints. In so doing, we also check whether these models can
account for a sizably negative running of the spectral index, as
(weakly) favored by current data. We focus on models where the cosmic
expansion was driven by a single, self--interacting scalar inflaton
field $\phi$. Moreover, we only consider models with simple potentials, which have been
suggested for reasons not related to PBH formation.

This paper is organized as follows: In section 2 we present a brief
review of the Press-Schechter formalism \cite{Press-Schechter}
describing PBH formation. In section 3 we first summarize the current
bounds on the observational parameters, and review their calculation
from the potential of the inflaton field. We then systematically analyze
three small--field and five large--field models. In section 4 we
present our conclusions.

\section{Primordial Black Holes Formation}

PBHs are black holes that result from the collapse of density fluctuations
\cite{Zeldovich, Hawking1}. They are very sensitive cosmological probes for
physics phenomena occurring in the early universe. They could be formed by
many different mechanisms, \textit{e.g.}, from initial density inhomogeneities
\cite{Zeldovich}, a softening of the equation of state \cite{khlopov} (\textit{e.g.} in
phase transition or during the preheating period after inflation), collapse of
cosmic string loops \cite{Hawking2}, bubble collisions \cite{bubble}, collapse
of domain walls \cite{domain walls}, etc.

The idea that large amplitude matter overdensities in the early
universe could have collapsed through self--gravity to form PBHs was
first studied by Zel'dovich and Novikov \cite{Zeldovich}, and then by
Hawking \cite{Hawking1}. This theory suggests that large amplitude
inhomogeneities in the early universe overcome internal pressure
forces and collapse to form black holes. A lower threshold for the
amplitude of such homogeneities
$\delta_{\text{th}}\equiv(\delta\rho/\rho)_{\text{th}}$, was first
provided by Carr \cite{Carr2}, giving $\delta_{\text{th}}\approx1/3$
at the time of radiation domination (RD).\footnote{Niemeyer and
  Jedamzik \cite{Jedamzik} carried out numerical simulations and found
  the threshold for PBH formation to be $0.7$. We have shown
  \cite{encieh} that PBHs abundance is sensitive to the value of
  $\delta_{\rm th}$.}  The probability of PBH formation is a useful
tool to constrain the mean amplitude of inhomogeneities on scales
which cannot be probed by any other method. Since PBHs behave like
matter, their contribution to the energy density increases with time
during the RD epoch. For this reason, the PBHs formed considerably
before the end of RD are the most relevant to cosmology. We have
focused in our study \cite{encieh} on these kind of PBHs and we also
considered the standard case of PBHs formation, which applies to
scales which have left the horizon at the end of
inflation.\footnote{It has been shown \cite{earlyPBH} that PBHs can
  also form on scales which never leave the horizon during inflation,
  and therefore never become classical. We do not consider this
  contribution.} We only consider Gaussian and spherically symmetric
perturbations and we assume that the mass of the PBH formed is
proportional to the mass of the horizon mass at horizon entry,
$M_{\text{PBH}}=\gamma M_{\text{PH}}$.\footnote{A simple analytical
  calculation suggest that $\gamma\simeq w^{3/2}\simeq 0.2$ during the
  radiation era \cite{Carr2}.}

Generally, for Gaussian primordial fluctuations, the probability density
$P(\delta;R)$, where $\delta$ is the density contrast averaged over a sphere
of radius $R$, is given by 
\begin{equation} \label{pofr}
P(\delta;R) = \dfrac {1} {\sqrt{2\pi} \sigma_{\delta}(R)} \exp\left(
-\dfrac{\delta^{2}} {2\sigma_{\delta}^{2}(R)} \right)\, . 
\end{equation}
Here, the dispersion (mass variance) $\sigma_{\delta}(R)$ is computed using a
Gaussian window function $W(kR) = \exp\left(-\dfrac{k^{2}R^{2}}{2} \right)$: 
\begin{equation} \label{sigma}
\sigma_{\delta}^{2}(R) = \int_{0}^{\infty} W^{2}(kR)
\mathcal{P}_{\delta}(k)\dfrac{\text{d}k} {k}\, , 
\end{equation}
where $\mathcal{P}_{\delta}(k)$ is the power spectrum of $\delta$ which is
related to the power spectrum of curvature perturbations on comoving
hypersurfaces as follows \cite{Lyth Book}: 
\begin{equation} \label{pofk}
\mathcal{P}_{\delta}(k,t) = \dfrac{4(1+w)^2} {(5+3w)^2} \left(
\dfrac{k}{aH}\right)^4 \mathcal{P}_{\mathcal{R}_{c}}(k)\, . 
\end{equation}
Therefore the probability $f(\geq M)$ that a PBH with mass $\geq M$ is formed
on a scale $R$ when that scale reenters the Hubble radius, is given
by\footnote{We will show that in practice $P(\delta; R)$ is such a rapidly
  decreasing function of $\delta$ above $\delta_{\rm th}$ that the upper
  cutoff $\delta_{\rm cut}$ is not important.}
\begin{equation} \label{fofm}
f( \geq M) = 2\, \gamma \int _{\delta_{\rm th}}^{\delta_{\rm cut}} P(\delta;
M(R))\text{d}\delta\, . 
\end{equation}

In order to complete the calculation we just need to relate the PBH mass $M$
to the comoving smoothing scale $R$ when the scale enters the horizon,
$R=(aH)^{-1}$. It is straightforward to show that in the RD era 
\begin{equation} \label{R}
\dfrac {R} {1\ \text{Mpc}} = 5.54 \times 10^{-24} \gamma^{-\frac{1}{2}}
\left(\dfrac {M_{\text{PBH}}} {1\ \text{g}} \right)^{1/2} \left( \dfrac
     {g_{\ast}}{3.36} \right)^{1/6}\, ,
\end{equation}
where $g_{\ast}$ is the effective number of relativistic degrees of freedom
which is expected to be of order $100$ in the early universe. 

The power spectrum of primordial scalar fluctuations at scales $k\simeq k_R$
is typically parameterized as a power--law with power $n_S$, 
$\mathcal{P}_{\mathcal{R}_{c}}(k) = \mathcal{P}_{\mathcal{R}_{c}}
(k_R)(k/k_R)^{n_S(R)-1}$, with $k_R = 1/R$. With this ansatz, the variance of
the primordial density field at horizon crossing is given by 
\begin{equation} \label{sigmaR}
\sigma_{\delta}^2 (R) = \dfrac {2(1+w)^2} {(5+3w)^2}
\mathcal{P}_{\mathcal{R}_{c}}(k_R) \Gamma[(n_S(R)+3)/2] \, ,  
\end{equation}
for $n_S(R)>-3$.

In order to relate the scales relevant for PBH formation to the scales
probed by data on the CMB and LSS, we parameterize the power spectrum
as
\begin{equation} \label{eqn}
\mathcal{P}_{\mathcal{R}_c}(k_R) =
\mathcal{P}_{\mathcal{R}_c}(k_0)(k_R/k_0)^{n(R)-1}\, . 
\end{equation}
It is important to distinguish between $n_S(R)$ and $n(R)$ at this
point. $n_S(R)$ describes the {\em slope} of the power spectrum at scales
$k\sim k_R = 1/R$, whereas $n(R)$ fixes the {\em normalization} of the
spectrum at $k_R \gg k_0$. The two powers are identical if the spectral index
is strictly constant, \textit{i.e.} if neither $n_S$ nor $n$ depend on $R$. However, in
this case CMB data imply \cite{WMAP7} that $n = n_S$ is close to
unity. Eqs.(\ref{sigmaR}) and (\ref{eqn}) then give a very small variance,
leading to essentially no PBH formation. 

Significant PBH formation can therefore only occur in scenarios with running
spectral index \cite{Josan}. We parameterize the scale dependence of $n$ as
\cite{Kosowsky}:
\begin{equation} \label{nofr}
 n(R) \quad = \quad n_S (k_0) - \frac{1}{2!} \, \alpha_S\, \ln \left( k_0 R
 \right) + \dfrac {1} {3!} \, \beta_S\, \ln^2 \left( k_0 R \right) +\dots\, ; 
\end{equation}
recall that we are interested in $R \ll 1/k_0$, \textit{i.e.} $\ln(k_0 R) < 0$.
The parameters $\alpha_S$ and $\beta_S$ denote the running of the effective
spectral index $n_S$ and the running of the running, respectively: 
\begin{eqnarray} \label{runpar}
n_S (k_0) \quad &\equiv& \quad \left. \dfrac {d\ln\mathcal{P}_{\mathcal{R}_c}}
{d \ln k} \right|_{k=k_0}\, , 
\nonumber \\
\alpha_S (k_0) \quad & \equiv &\quad \left. \dfrac {d n_S} {d\ln
  k}\right|_{k=k_0} \, , 
\nonumber \\
\beta_S (k_0) \quad & \equiv & \quad \left. \dfrac {d^2 n_S} {d\ln^2
  k}\right|_{k=k_0}\, . 
\end{eqnarray}
Eq.(\ref{nofr}) illustrates the difference between $n(R)$ and $n_S(R)$. The
latter has an expansion similar to eq.(\ref{nofr}), but with the usual
Taylor--expansion coefficients, $1$ in front of $\alpha_S$ and $1/2$ in front
of $\beta_S$. One therefore has 
\begin{equation} \label{nsofn}
n_S(R)\quad = \quad n(R) - \frac {1}{2} \, \alpha_S\, \ln \left( k_0 R \right)
+ \frac{1}{3} \, \beta_S \, \ln^2 \left( k_0 R \right) + \dots \, . 
\end{equation}
Setting $n_S(k_0)=1$ for simplicity, eq.(\ref{nsofn}) implies $n_S(R) = 2 n(R)
- 1$ for $\beta_S = 0$, and $n_S(R) = 3 n(R) - 2$ for $\alpha_S = 0$. We will
compute the variance $\sigma(R)$, and hence the PBH fraction, for the former
relation. 

\begin{figure}[h!]
\centering
\includegraphics[width=1\textwidth]{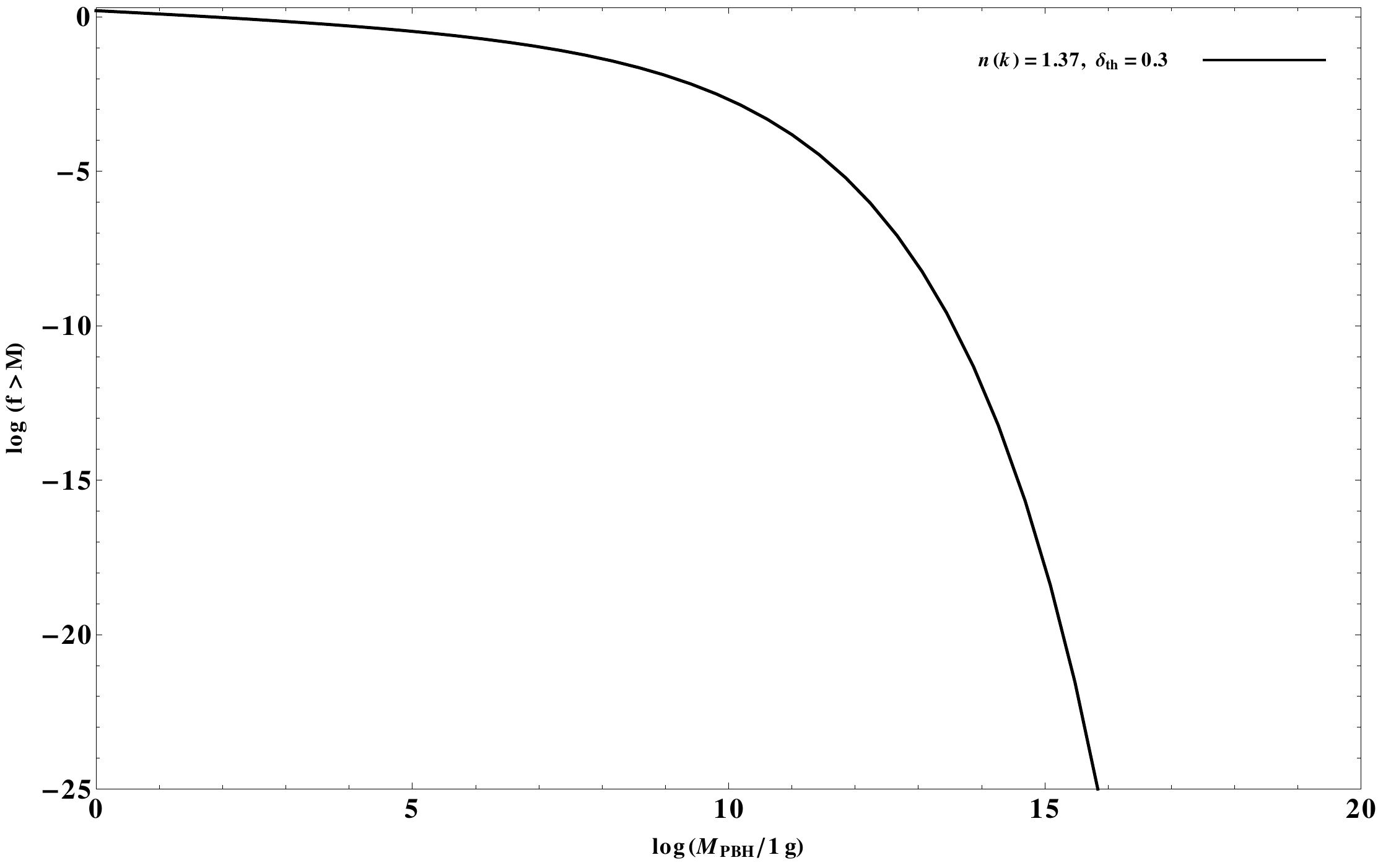}
\caption{Fraction of the energy density of the universe collapsing into PBHs
  as a function of the PBH mass.} 
\label{fig:PBH}
\end{figure}

The result of this calculation is shown in figure~\ref{fig:PBH}. Here
we have fixed $\gamma = 0.2$, and show results for the threshold
$\delta_{\rm th}=1/3$. Due to Hawking radiation \cite{Hawking3}, PBHs
contributing to Dark Matter (DM) today must have $M_{\rm PBH} \gsim
10^{15}$ g; at this mass, they saturate the DM relic density if
$f\simeq 5\times10^{-19}$ \cite{observation}. Figure~\ref{fig:PBH}
shows that this requires $n(R) \simeq 1.37$ for $\delta_{\rm th} =
0.3$. In order to get long--lived PBHs the amplitude of the
perturbations at PBH scales must therefore exceed that at CMB scales
by a factor $10^3$--$10^4$. Current data favor a negative or at best
slightly positive value of $\alpha_S$ at the CMB pivot scale, as well
as a spectral index at the pivot scale somewhat below $1$. The first
two terms in eq.(\ref{nofr}) can thus not lead to PBH formation; so in
the next section we study single--field inflation models by also
considering the ``running of the running of the spectral index'' for
PBHs formation. When possible, we will compare this to an exact
calculation of the power of density perturbation at scales relevant
for PBH formation.

\section{Inflation Models and Primordial Black Holes}

Most models of inflation predict an approximately scale--free spectrum
with a spectral index $n_S$ (as well as $n$) close to the
scale--invariant (Harrison--Zel'dovich) case $n_S = n = 1$. As shown,
a significant number of long--lived PBHs can only be produced for $n >
1$ (a ``blue spectrum''), since these values lead to more power on
small scales \cite{PBH_formation}.  Observational limits (both from
Hawking radiation and the fact that PBHs must not overclose the
universe) strongly constrain $n_S$ \cite{pbh_constraint}. This,
therefore, yields a constraint on inflationary models that is
independent of the cosmological constraints from the CMB and
LSS. However, we will see below that in most simple models of
inflation constraints on model parameters derived from the latter are
far more stringent than the PBH constraint, to the point of making the
formation of long--lived PBHs impossible.

A given inflation model can be described to lowest order in slow roll
by three independent parameters: the normalization of the curvature
perturbation spectrum $\mathcal{P}_{\mathcal{R}_c}$ at some scale, the
tensor--to--scalar ratio $r$, and the spectral index $n_S$. If we wish
to include higher-order effects, we have the forth and fifth
parameters describing the running $\alpha_S$, and the running of
running of the spectral index $\beta_S$.

Observational bounds on $\mathcal{P}_{\mathcal{R}_c}$, $r$, $n_S$ and
$\alpha_S$ at the pivot $k_{\text{pivot}}=0.015\, \text{Mpc}^{-1}$, where
$n_S$ and $\alpha_S$ are essentially uncorrelated, are reported in
\cite{SPT} as follows:\footnote{The amplitude of the primordial scalar
  fluctuations is reported at the ``COBE'' scale $k_0=0.002 \,
  \text{Mpc}^{-1}$.} 
\begin{eqnarray} \label{range}
n_S(k_{\text{pivot}}) & = & 0.9751 \pm 0.0110 \, , \nonumber \\
\alpha_S(k_{\text{pivot}}) & = & -0.017 \pm 0.012\, , \nonumber \\
\mathcal{P}_{\mathcal{R}_c}(k_0) & = & (2.33 \pm 0.092 )\times 10^{-9}\, ,
\nonumber \\ 
r & < & 0.17\, (95\% \text{CL})\, . 
\end{eqnarray}
By requiring $n_S(k)\in \left[0.9531, 0.9971 \right]$ and $\alpha_S\in\left[-0.041, 0.007 \right]$ for all $k\in\left[10^{-4}, 10 \right] $ Mpc$^{-1}$ (\textit{i.e.} down to the Lyman--$\alpha$ range), we find from eq.(\ref{nofr}) that values of $\beta_S$ up to $0.017$ are allowed. The ranges of $n_S$ and $\alpha_S$ are simply in $2\,\sigma$ range of (\ref{range}), and the range of $k$ encompasses all cosmologically relevant scales. Here we used the error bars derived from the
analysis of the ``SPT+WMAP7+BAO+$H_0$+\textit{Clusters}'' data
set.\footnote{We ignore possible tensor modes, which is appropriate
  for small--field models. Allowing a sizable contribution from tensor
  modes changes the mean value of $\alpha_S$ \cite{WMAP7}, but
  unfortunately the pivot scale where the spectral index and its
  running are uncorrelated is not reported in running+tensor model. In
  the SPT data, the inflation parameters are not reported in
  running+tensor model. So although in the large--field models, tensor
  modes are not negligible we will assume that the upper bound of
  $\beta_S$ is the same as in small--field models. Note that the
  precise value of the upper bound on $\beta_S$ derived here is not
  important for our analysis, since it is in any case well above the
  lower bound needed for successful PBH formation.} This upper bound on $\beta_S$
is lower than the bound found from the ``WMAP+$H_0$+BAO'' data set
\cite{encieh}, since the older data set allowed somewhat larger values
of $\alpha_S$. Note that the estimates for spectral index and its
running are highly correlated for the typical COBE scale, $k_0=0.002\,
\text{Mpc}^{-1}$ \cite{correlation}.

Eq.(\ref{nofr}) shows that for $\alpha_S = 0$ we only need
$\beta_S(k_0) \simeq 0.0015$ in order to generate sufficiently large
density perturbations to allow formation of $10^{15}$ g PBHs. Even if
we set $\alpha_S(k_0)$ equal to its central value, $\alpha_S(k_0) =
-0.017$ , we only need $\beta_S(k_0) \simeq 0.0028$. In this
model--independent analysis including the running of the running of
the spectral index thus easily allows to accommodate PBH formation in
scenarios that reproduce all current cosmological observations at
large scales.

Of course, this kind of model--independent analysis does not show
whether simple, reasonably well--motivated inflationary models exist
that can generate a sufficiently large $\beta_S$. In the following we
study different models of inflation and check whether they can lead to
PBHs formation. As a by--product, we also check whether these models
can accommodate a sizably negative value of $\alpha_S$, as indicated
by current data.

In order to calculate the spectral parameters $n_S, \, \alpha_S$ and $\beta_S$
defined in eqs.(\ref{runpar}), we need the first four slow--roll parameters,
defined as \cite{Lyth Book}:
\begin{eqnarray} \label{slow-roll}
\epsilon & \equiv& \dfrac {M_{\rm P}^2} {2} \left( \dfrac {V^{\prime}}
         {V}\right)^2 \, , 
\nonumber \\
\eta & \equiv & M_{\rm P}^2 \dfrac {V^{\prime\prime}} {V}\, ,
\nonumber \\
\xi^2 & \equiv & M_{\rm P}^4 \dfrac {V^{\prime} V^{\prime\prime\prime}} {V^2}\, ,
\nonumber \\
\sigma^3 & \equiv& M_{\rm P}^6 \dfrac {V^{\prime2}
  V^{\prime\prime\prime\prime}}{V^3}=2 M_{\rm P}^4\epsilon
\dfrac{V^{\prime\prime\prime\prime}}{V}\, . 
\end{eqnarray}
Here $V(\phi)$ is the inflaton potential, $M_{\rm P}$ is the reduced Planck
mass, and primes denote derivatives with respect to $\phi$. Note that
$\epsilon$ is positive by definition whereas in spite of the square, $\xi^2$
can be either positive or negative. The square in $\xi^2$ and cube in
$\sigma^3$ are to indicate that they are second and third--order in the
slow--roll expansion, respectively. These parameters must be less than one for
the slow--roll expansion to be valid. All these parameters are in general
scale--dependent, \textit{i.e.} they have to be evaluated at the value of
$\phi$ that the inflaton field had when the scale $k$ crossed out of the
horizon. The spectral parameters are related to these slow--roll parameters by
\cite{Lyth Book,running of running}:
\begin{eqnarray} \label{nab}
n_S & = & 1 - 6 \epsilon + 2 \eta\, ,
\nonumber \\
\alpha_S & = & -24 \epsilon^2 + 16 \epsilon \eta - 2\xi^2 \, ,
\nonumber \\
\beta_S & = & -192 \epsilon^3 + 192 \epsilon^2 \eta - 32 \epsilon \eta^2
-24\epsilon \xi^2 +2 \eta \xi^2 +2 \sigma^3 \, . 
\end{eqnarray}
In most (small--field) inflation models, eqs.(\ref{slow-roll}) imply two
strong inequalities between (combinations of) slow-roll parameters
(hierarchy):
\begin{eqnarray} \label{hierarchies}
|\epsilon| &\ll& |\eta|\, ,
\nonumber \\
|\epsilon \eta| & \ll & |\xi^2|\, .
\end{eqnarray}
The first relation means that $n_S-1$ is essentially determined by
$\eta$. Similarly, both relations together imply that $\alpha_S$ is basically
fixed by $\xi^2$, while only the last two terms in the expression for
$\beta_S$ are relevant; these two terms are generically of similar order of
magnitude. 

Along with these, another crucial inflationary observable is the influence of
gravitational waves, relative to density perturbations, on large--angle
microwave background anisotropies, given by \cite{Lyth Book}
\begin{equation} \label{tensor}
 r\equiv\dfrac{C_2(\text{grav})}{C_2(\text{dens})}\simeq 14\epsilon \, .
\end{equation}
Combining these equations gives \cite{tensor}
\begin{eqnarray} \label{tensor2}
\alpha_S & \simeq & 6\left( \dfrac{r}{7}\right) ^2+4\left( \dfrac{r}{7}\right)
(n_S-1)-2\xi^2\, , 
 \nonumber\\
\beta_S & \simeq & -15\left( \dfrac{r}{7}\right)^3-15\left(
\dfrac{r}{7}\right)^2(n_S-1)-2\left( \dfrac{r}{7}\right)(n_S-1)^2 
\nonumber \\
& + & \dfrac{\alpha_S}{2}\left[9\left( \dfrac{r}{7}\right)-(n_S-1)
  \right]+2\sigma^3 \, . 
\end{eqnarray}

\begin{figure}[h!]
\centering{\includegraphics[width=0.454\textwidth]{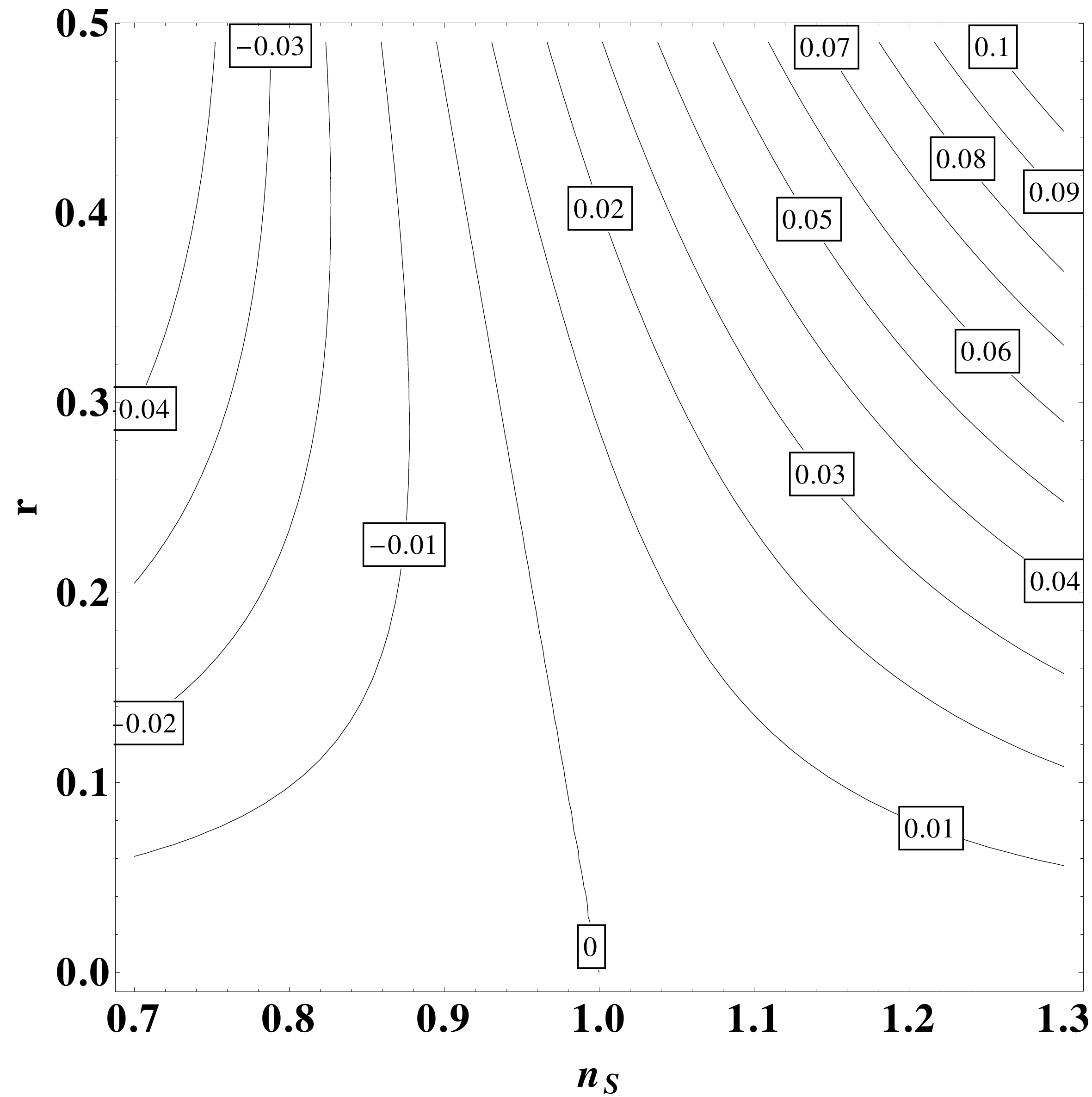}
\hspace{0.3cm}
\includegraphics[width=0.474\textwidth]{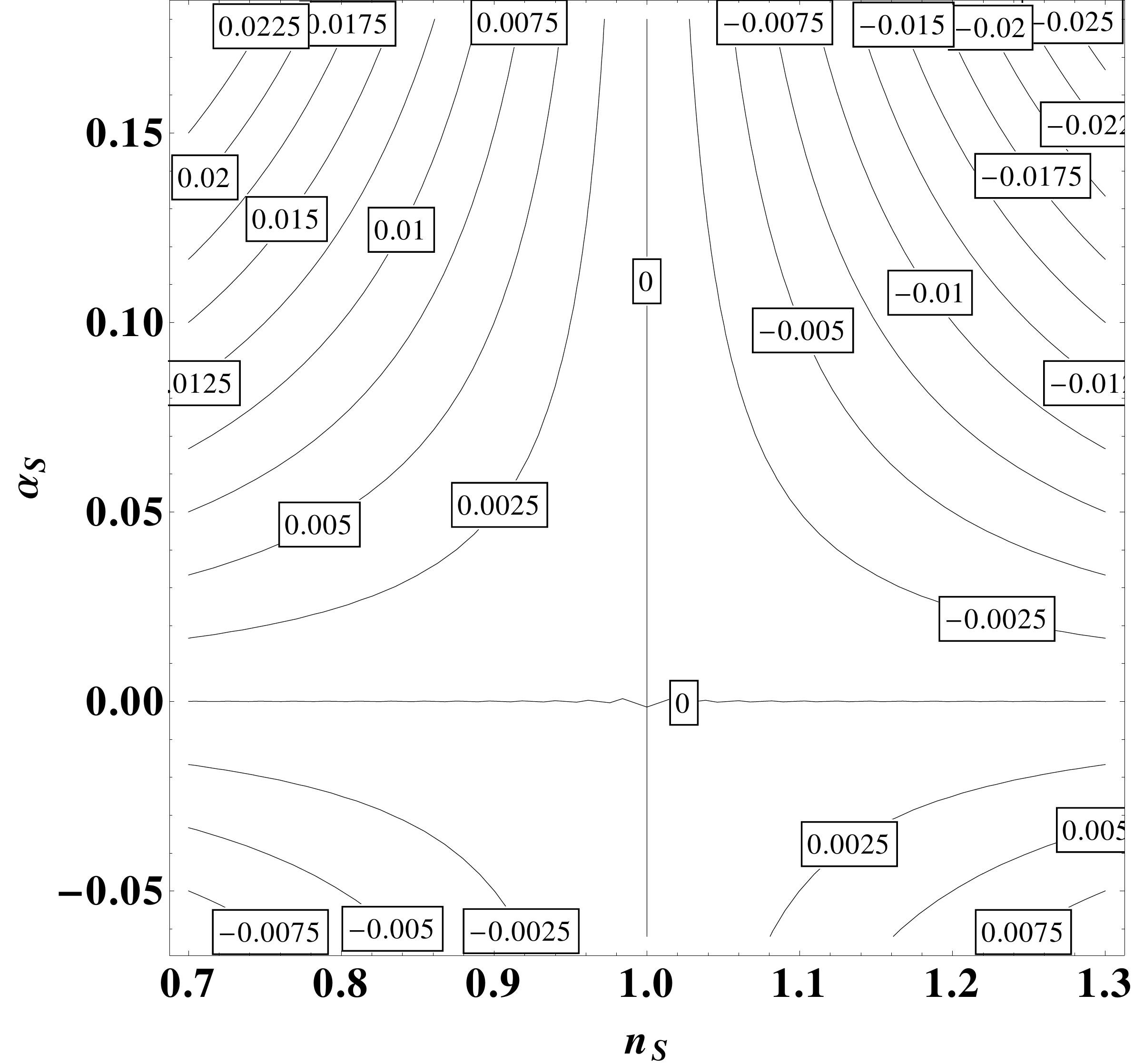}}
\caption{Contours of $\alpha_S + 2\xi^2$ (left) and of $\beta_S -
  2\sigma^3$ (right); the right frame assumes negligible tensor modes,
  $r = 0$.}
\label{fig:contour}
\end{figure}

In the left frame of figure~\ref{fig:contour} we show contours of
$\alpha_S + 2\xi^2$ in the $(n_s, r)$ plane, while contours of
$\beta_S - 2\sigma^3$ in the $(n_S, \alpha_S)$ plane are shown in the
right frame, assuming a negligible tensor--to--scalar ratio. We see
that the observational constraints on $n_S$ and $r$ imply that
$\alpha_S+2\xi^2$ is very small, roughly $-9\times10^{-4}\leq
\alpha_S+2\xi^2\leq2.2\times10^{-3}$ if $n_S$ and $r$ are within their
current $2\,\sigma$ intervals. Any significant running must therefore
be due to $\xi^2$ \cite{running}. Similarly,
$-5\times10^{-4}\leq\beta_S-2\sigma^3\leq -4\times10^{-5}$ if $r=0$
and $n_S$ and $\alpha_S$ are within their $1\,\sigma$ intervals. Even
using $2\,\sigma$ intervals and allowing $r\leq0.17$, this range only
expands to $-0.0053\leq\beta_S-2\sigma^3\leq 0.001$, so that
significant positive running of the running can only be due to
$\sigma^3$.

In slow--roll approximation, the absolute normalization of the power spectrum
is given by
\begin{equation} \label{power}
\mathcal{P}_{\mathcal{R}_{c}} = \dfrac {1} {12\pi^2 M_{\rm P}^6}
\dfrac{V^3}{V^{\prime2}} \, . 
\end{equation}

Finally, the number of $e$--folds of slow--roll inflation that occurred from
the time $t_{\ast}$ when observable CMB scales first crossed the Hubble radius
during inflation to the epoch $t_{\text{end}}$ when inflation ended is given
by
\begin{equation} \label{efold}
N = \dfrac {1} {M_{\text{P}}^2} \int_{\phi_{\text{end}}}^{\phi_{\ast}} \dfrac
{V} {V^{\prime}} d\phi\, . 
\end{equation}
where $\phi_{\text{end}}$ is defined by ${\rm max}\left[
  \epsilon(\phi_{\text{end}}),|\eta(\phi_{\rm end})|\right] =1$; note that
inflation might end through dynamics of other fields coupled to the inflaton,
as in hybrid inflation. The observationally required value of $N$ depends
logarithmically on the reheating temperature.\footnote{Instantaneous reheating
  gives the minimum number of $e$--folds as one looks backwards to the time of
  perturbation production, while a prolonged period of reheating gives a
  larger number of $e$--folds.}  Assuming instantaneous change from inflation
to relativistic matter domination, a reasonable range of values of the number
of $e$--folds between $t_{\ast}$ and $t_{\text{end}}$ is taken to be
$N=54\pm7$ \cite{e-folds}. Requiring baryogenesis to take place at or above
the electroweak scale implies that $N\gtrsim30$. A value of $N\simeq60$
corresponds to a GUT scale reheating.\footnote{Arbitrarily many $e-$folds of
  inflation might have occurred at $t < t_\ast$, as in ``eternal''
  inflation. $N$ of eq.(\ref{efold}) is a lower bound on the total number of
  $e-$folds of inflation.}

In the following we study the possibility of PBH formation in two different
categories of inflation models: small--field models and large--field
models. Hybrid models \cite{Hybrid Linde} are not studied here because in
these models, PBH formation can occur by different mechanisms. (For analyses
of PBH formation in hybrid models, see \cite{PBHs hybrid}.) As noted above,
the spectral index will have to increase at very small scales (very large $k$)
in order to allow PBH formation, but we will also check whether the models we
analyze are compatible with a sizably negative value of $\alpha_S$ at scales
probed by the CMB and LSS data, as indicated by eqs.(\ref{range}).

\subsection{Small--field models}

Small--field models are defined as those for which the variation in the
inflaton field is less than the reduced Planck mass. Typically, $\epsilon$ and
hence the amplitude of gravitational waves generated in such models is small
and the spectral index and its running provide the key observational
discriminators.

\subsubsection{Hilltop/inflection point inflation}

A popular ansatz for the small field inflaton potential is given by \cite{review}
\begin{equation}\label{small field}
 V(\phi)=V_0\left[1-\left(\dfrac{\phi}{\mu} \right)^p  \right] \, ,
\end{equation}
where $V_0,\, \mu$ and $p$ are positive constants.\footnote{This
  potential is unbounded from belove for $\phi \rightarrow
  \infty$. There must be additional terms that prevent this. Here we
  follow the usual assumption that these terms do not affect the
  dynamics of inflation.} This potential is equivalent to the
potential $V(\phi) = \Lambda^4 - \lambda M_{\text{P}}^{4-n}
\dfrac{\phi^n} {n}$ in the literature \cite{Hilltop inflation} which can be
specialized to several distinct models: \textit{e.g.}  hilltop ($n=2$
or $n=4$) and inflection point ($n=3$). We consider the case that the
dominant term is the leading one, $V_0$. When $p$ is an integer and
greater than $2$, such a potential may be generated by the
self--coupling of the inflaton at tree--level.

For $p>0$, the hierarchies (\ref{hierarchies}) hold among slow--roll
parameters. So the spectral parameters are given by:
\begin{eqnarray} \label{model1}
n_S-1&\simeq&-2p(p-1)\left( \dfrac{M_{\text{P}}}{\mu}\right)^2
\left(\dfrac{\phi}{\mu} \right)^{p-2}\, ; \nonumber \\ 
\alpha_S&\simeq&-2p^2(p-1)(p-2)\left(\dfrac{M_{\text{P}}}{\mu} \right)^4
\left( \dfrac{\phi}{\mu}\right)^{2(p-2)}\, ; \nonumber\\ 
\beta_S&\simeq&-4p^3(p-1)(p-2)^2\left( \dfrac{M_{\text{P}}}{\mu}\right)^6
\left(\dfrac{\phi}{\mu} \right)^{3(p-2)}\, , 
\end{eqnarray}
Inflation ends at $\phi_{\text{end}}\lesssim\mu$, and in order to have a small
field model we take $\mu\lesssim M_{\text{P}}$. Then
\begin{equation}\label{N1}
 N=-\dfrac{p-1}{p-2}+\dfrac{1}{p(p-2)}\left(\dfrac{\mu}{M_{\text{P}}}
 \right)^2 \left(\dfrac{\mu}{\phi} \right)^{p-2}\, . 
\end{equation}
For $p>2$, the first term in eq.(\ref{N1}) can be neglected. We then find,
independently of $\mu$:
\begin{eqnarray} \label{model11}
n_S-1&\simeq&- \dfrac{p-1}{p-2}\dfrac{2}{N}\, , \nonumber \\
\alpha_S&\simeq&-\dfrac{p-1}{p-2}\dfrac{2}{N^2}=\dfrac{1}{N}(n_S-1)\, ,
\nonumber\\ 
\beta_S&\simeq&-\dfrac{p-1}{p-2}\dfrac{4}{N^3}=\dfrac{2}{N^2}(n_S-1)\, .
\end{eqnarray}

\begin{figure}[h!]
\centering
\includegraphics[width=1.\textwidth]{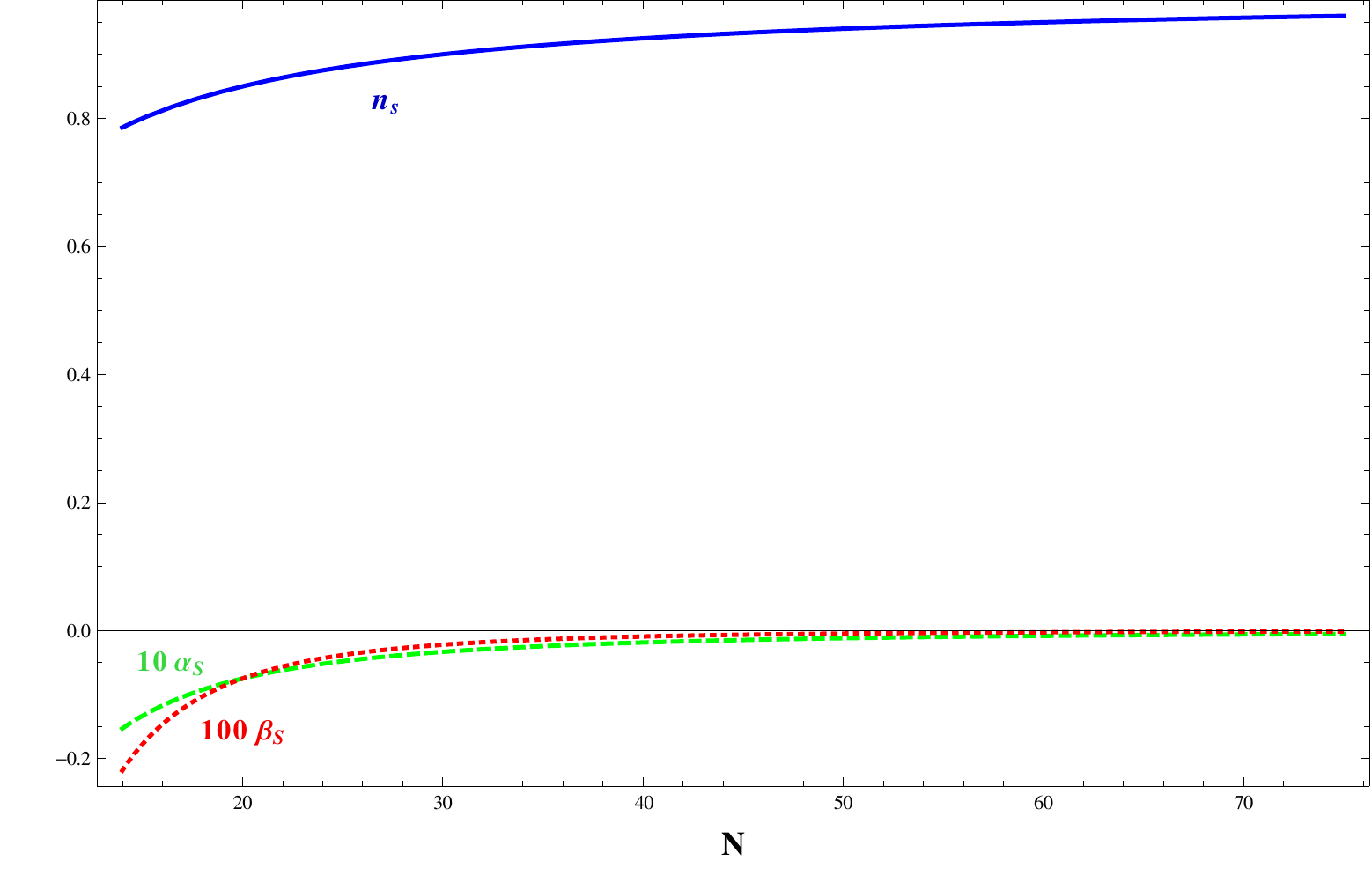}
\caption{Illustrating the dependence according to eqs.(\ref{model11}) of $n_S$
  (solid curve), $10 \alpha_S$ (dashed curve) and $100 \beta_S$ (dotted
  curve), on the number of $e$--folds before the end of inflation, for the
  fixed value of $p=4$.}
\label{fig:sf1}
\end{figure}

In figure \ref{fig:sf1} the spectral index, its running and its running of
running are shown as functions of the number of $e$--folds before the end of
inflation, for $p=4$. 

It is clear that in this model both $n_S-1$ and $\alpha_S$ are negative, but
it is not possible to reproduce the observed central value of $\alpha_S$,
which would require $\alpha_S \sim {\cal O}(n_S-1)$ (see figure
\ref{fig:sf1}). Moreover, the value of $\beta_S$ is also negative. So the
conclusion is that this model cannot produce sufficient high density
fluctuations at small scales to produce PBHs.

In the case at hand, the power spectrum can be calculated exactly as function
of $N$. Again neglecting the first term in eq.(\ref{N1}) we have from
eq.(\ref{power}):
\begin{equation}
\mathcal{P}_{\mathcal{R}_c} (N) = \dfrac{1}{12\pi^2} \dfrac{V_0 \mu^2}{p^2 M_{\rm
    P}^6} \left[ N p(p-2) \dfrac{M_{\rm P}^2}{\mu^2}
  \right]^{\frac{2p-2}{p-2}}\, .
\end{equation}
Note that the exponent is positive for $p>2$. This implies less power at
smaller $N$, \textit{i.e.} at smaller length scales.

\subsubsection{Running mass inflation}

Another small--field model of interest is the running--mass model
\cite{Stewart}. The possibility of PBHs formation in this model is studied in
detail in \cite{encieh}. This model is based on the inflationary potential
\begin{equation} \label{rm1}
V(\phi) = V_0 + \dfrac {1} {2} m_\phi^2(\phi) \phi^2\, ,
\end{equation}
where $\phi$ is a real scalar. The potential (\ref{rm1}) by itself would lead
to eternal inflation, so again some terms need to be added, and we again
assume that these terms do not affect the dynamics of the inflaton during the
slow--roll phase. During inflation, the potential is dominated by the constant
term $V_0$. Here $m_\phi^2(\phi)$ is obtained by integrating a
renormalization group equation of the form
\begin{equation} \label{rm2}
\dfrac {d m_\phi^2} {d\ln \phi} \equiv \beta_m = - \dfrac {2C} {\pi} \alpha\,
\widetilde{m}^2 + \dfrac {D} {16\pi^{2}}|\lambda_Y|^2 m_s^2\, , 
\end{equation}
where $\beta_m$ is the $\beta-$function of the inflaton mass parameter which
arises from the gauge interaction with coupling $\alpha$ and from the Yukawa
interaction $\lambda_Y$.  $C$ and $D$ are positive numbers of order one, which
depend on the representations of the fields coupling to $\phi$,
$\widetilde{m}$ is a gaugino mass parameter, while $m_s^2$ is the scalar SUSY
breaking mass--squared of the scalar particles interacting with the inflaton
via Yukawa interaction $\lambda_Y$.

Over a sufficiently small range of $\phi$, or small inflaton coupling, we can
perform a Taylor expansion around $\phi_*$
\begin{equation} \label{rm3}
V = V_0 + \dfrac {1} {2} m_\phi^2 (\phi_*) \phi^2 + \dfrac {1} {2}
\ c\ \phi^2\ \ln \left( \dfrac {\phi} {\phi_*} \right) + \dfrac {1}
         {4}\ g\ \phi^2\ \ln^2 \left( \dfrac {\phi} {\phi_*} \right)\, ,
\end{equation}
where $\phi_*$ is the local extremum of the potential. Here $c\equiv
\left. \dfrac{d m_\phi^2}{d\ln \phi} \right|_{\phi=\phi_*}$ is given
by the $\beta-$function, and $g \equiv\left. \dfrac{d^2 m_\phi^2} {d
    (\ln \phi)^2} \right|_{\phi=\phi_*}$ is given by the scale
dependence of the parameters. By having the potential in hand and
noting that the hierarchies (\ref{hierarchies}) among slow--roll
parameters hold in this model, we find the spectral parameters,
\begin{eqnarray} \label{rm4}
n_S - 1 & = & 2 \frac{c M_{\rm P}^2} {V_0} \left[ L + 1 + \frac {g}{2c} \left(
  L^2 + 3 L + 1 \right) \right]\, , 
\nonumber \\
\alpha_S & = & -2 \left( \frac {c M_{\rm P}^2}{V_0} \right)^2 L \left[ 1 +
  \frac{g} {2c} \left( 2L + 3 \right) \right] \left[ 1 + \frac {g} {2c}
  \left(L + 1 \right) \right] \, ,\\ 
\beta_S & = & 2 \left( \frac {c M_{\rm P}^2}{V_0} \right)^3 L \left[ 1 + \frac
  {g}{2c} \left( L+1 \right) \right] \left[ 1 + \frac {g}{2c} \left( 3L+2
  \right) + \frac {g^2} {2c^2} \left( 3 L^2 + 5L + \frac{3}{2} \right) \right] 
\nonumber \, ,
\end{eqnarray}
where $L \equiv \ln \dfrac{\phi}{\phi_*}$. Clearly the spectral index is
not scale--invariant unless $c$ and $g$ are very close to zero.

An important feature of the running--mass model is that because of a
consistency relation among the spectral index and its running, it is not
possible \cite{encieh} to get large negative running, which is presently
favored by observation:
\begin{equation} \label{rm5}
\alpha_S \geq - \frac { \left(n_S - 1 \right)^2} {4}\, .
\end{equation}

On the other hand, we showed in \cite{encieh} that over a narrow region of
allowed parameter space this model predicts sufficiently large density
perturbations at small scales to allow $10^{15}$ g PBHs to saturate the Dark
Matter density.

\subsubsection{Inverse power law inflation}

A generic feature of models in nonperturbative gauge dynamics in SUSY
\cite{dynamical SUSY} is the presence of scalar potentials of the form
$\dfrac{\Lambda_3^{p+4}}{\phi^p}$, where the index $p$ and the scale
$\Lambda_3$ depend on the underlying gauge group. Like models of hybrid
inflation \cite{Hybrid Linde}, these models are characterized by a potential
dominated by the constant term $V_0$ and require coupling to another sector to
end inflation when $\phi$ reaches the critical value $\phi_c$. Unlike standard
hybrid inflation models, models of this type postulate a field far from the
minimum of the potential.

We take the potential to be described by a single degree of freedom $\phi$, of
the general form
\begin{equation}\label{dsi}
V(\phi)=V_0+\dfrac{\Lambda_3^{p+4}}{\phi^p}+...\, ,
\end{equation}
where the dots represent nonrenormalizable terms suppressed by powers
of the Planck mass, which are not relevant for the present discussion,
but will prevent $\phi$ from ``running away'' to infinity. In the
limit $\phi\ll\left\langle \phi\right\rangle $, the term
$\sim\phi^{-p}$ dominates the dynamics:
\begin{eqnarray} \label{dsi1}
 V(\phi) & \simeq & V_0+\dfrac{\Lambda_3^{p+4}}{\phi^p},\, \, \, \, \, \,
 \phi\ll\left\langle \phi\right\rangle 
 \nonumber\\
&=&  V_0\left[ 1+\alpha\left(\dfrac{M_{\text{P}}}{\phi} \right)^p \right]\, ,
\end{eqnarray}
where $\alpha\equiv\dfrac{\Lambda_3^{p+4}}{M_{\text{P}}^{p}V_0}$. We
assume that the constant $V_0$ dominates the potential, or $\alpha \ll
(\phi/M_{\text{P}})^p$. In this case also hierarchies
(\ref{hierarchies}) hold among the slow--roll parameters which leads
to the following spectral parameters:
\begin{eqnarray} \label{dsi2}
n_S-1 & \simeq & 2\, p(p+1)\, \alpha \left( \dfrac{M_{\text{P}}}{\phi}
\right)^{p+2}\, ;  
 \nonumber \\
\alpha_S & \simeq & -2\, p^2(p+1)(p+2)\, \alpha^2 \left(
\dfrac{M_{\text{P}}}{\phi}\right)^{2(p+2)}\, ; 
 \nonumber\\
\beta_S &\simeq & 4\, p^3(p+1)(p+2)^2\, \alpha^3 \left(
\dfrac{M_{\text{P}}}{\phi}\right)^{3(p+2)}\, .
\end{eqnarray}
The number of the $e$--folds $N$ is given by
\begin{eqnarray} \label{dsi3}
 N\simeq\dfrac{1}{p(p+2)\, \alpha}\left[\left(
   \dfrac{\phi_c}{M_{\text{P}}}\right)^{p+2}- \left(
   \dfrac{\phi}{M_{\text{P}}}\right)^{p+2}  \right]\, , 
\end{eqnarray}
where $\phi_c$ is the critical value at which inflation ends. The value of
$\phi_c$ is in general determined by a coupling of the field to some other
sector of the theory which we have here left unspecified. Note that from
eq.(\ref{dsi3}), for $\phi\ll\phi_c$ the number of $e$--folds approaches a
constant, which we call $N_{\text{tot}}$,
\begin{eqnarray} \label{dsi4}
N_{\text{tot}}\equiv \dfrac{1}{p(p+2)\, \alpha}\left(
\dfrac{\phi_c}{M_{\text{P}}}\right)^{p+2}\, . 
\end{eqnarray}
This puts an upper limit on the total amount of expansion that takes place
during the inflationary phase, although that upper bound can in principle be
very large. Using eqs.(\ref{dsi2})--(\ref{dsi4}), we can rewrite the
cosmological parameters as functions of the number $N$ of $e-$folds before the
end of inflation:
\begin{eqnarray} \label{dsi5}
n_S-1 & \simeq &
\dfrac{p+1}{p+2}\dfrac{2}{N_{\text{tot}}\left(1-\dfrac{N}{N_{\text{tot}}}
  \right) }\, , 
 \nonumber \\
\alpha_S & \simeq &
-\dfrac{p+1}{p+2}\dfrac{2}{{N_{\text{tot}}^2}\left(1-\dfrac{N}{N_{\text{tot}}}
  \right)^2 }=-\dfrac{p+2}{p+1}\dfrac{(n_S-1)^2}{2}\, , 
 \nonumber\\
\beta_S &\simeq &
\dfrac{p+1}{p+2}\dfrac{4}{{N_{\text{tot}}^3}\left(1-\dfrac{N}{N_{\text{tot}}}
  \right)^3 }=\left( \dfrac{p+2}{p+1}\right)^2 \dfrac{(n_S-1)^3}{2}\, . 
\end{eqnarray}
\begin{figure}[h!]
\centering
\includegraphics[width=1.\textwidth]{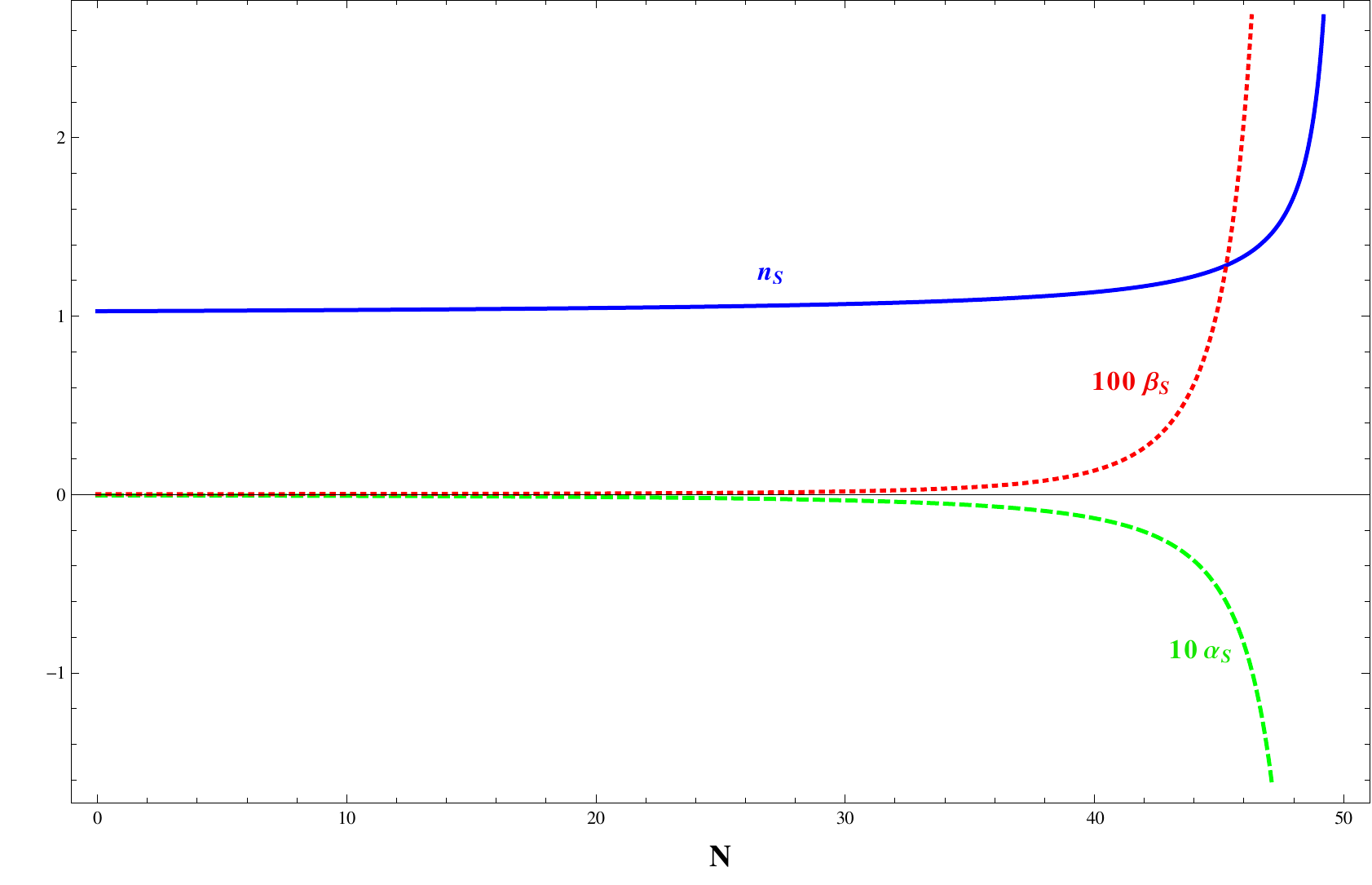}
\caption{Spectral parameters as a function of the number of $e$--folds
  $N\propto \ln(k)$ for $p=1$. Note especially the rapid approach to
  scale--invariance at short wavelengths (small $N$).} 
\label{fig:dsi}
\end{figure}

This model thus predicts $n_S>1$, which is currently disfavored at more than 2
standard deviation. Moreover, figure~\ref{fig:dsi} shows that the spectrum
becomes scale--invariant towards the end of inflation, \textit{i.e.} it
becomes less blue at smaller length scales, as also indicated by the negative
value of $\alpha_S$. In combination with the constraint that $|n_S-1|\ll1$ at
CMB scale, this implies that this model cannot accommodate PBH formation.

This can also be seen by directly computing the power spectrum as a function
on $N$:
\begin{equation} \label{dsi6}
\mathcal{P}_{\mathcal{R}_c}(N)=\dfrac{V_0}{12\pi^2M_{\rm
    P}^4}\dfrac{1}{\alpha^2p^2}\left[\alpha\, p(p+2)(N_{\rm tot}-N)
  \right]^{\frac{2p+2}{p+2}}\, ,
\end{equation}
where we have used (\ref{dsi3}). The power does increase with
decreasing $N$, but only by a small amount. For example, the ratio of
the power at the end of inflation $(N=0)$ to that at the COBE scale is
\begin{equation}\label{dsi7}
q \equiv \dfrac {\mathcal{P}_{\mathcal{R}_c}(N=0)}
{\mathcal{P}_{\mathcal{R}_c} (N_{\rm COBE})} = \left( \dfrac{N_{\rm
      tot} - N_{\rm COBE}} {N_{\rm tot}} \right)^{-\frac{2p+2}{p+2}}\, .
\end{equation}
On the other hand, the first eq.(\ref{dsi5}) gives 
\begin{equation}\label{dsi8}
n_S(N_{\rm COBE}) - 1 \simeq \dfrac {p+1} {p+2} \dfrac {2} { N_{\rm COBE}
  \left( \dfrac {N_{\rm tot}} {N_{\rm COBE}} -1 \right) }\, . 
\end{equation}
Eq.(\ref{dsi7}) can be rewritten as
\begin{equation} \label{dsi9}
\frac {1} { \dfrac{N_{\rm tot}}{N_{\rm COBE}} - 1 } = q^{\frac{p+2}{2p+2}}-1\, . 
\end{equation}
Inserting this into eq.(\ref{dsi8}) finally yields
\begin{equation}\label{dsi10}
q^{\frac{p+2}{2p+2}} = 1 + \dfrac{p+2}{2(p+1)} N_{\rm COBE} \left[ n_S(N_{\rm
    COBE}) - 1 \right] \simeq 2\, .  
\end{equation}
The power can therefore only increase by small amount in the course of
inflation; in contrast, PBH formation would require an increase by a factor
$10^7$ or so.

\subsection{Large--field models}

Large--field models are characterized by the condition $|\Delta\phi| \gtrsim
M_{\text{P}}$. Note that a super--Planckian field variation is a necessary
condition for the generation of an observable tensor-to-scalar ratio
\cite{gravitational waves}. On the other hand, such large field models raise
issues of stability in the presence of ``quantum gravity'' corrections, which
are suppressed by inverse powers of $M_{\rm P}$. These corrections should not
be important for small--field models, but need not be small for large--field
models.

\subsubsection{Power--law (a.k.a. chaotic) inflation}

The polynomial potential $V(\phi) = \Lambda^4 \left( \dfrac {\phi}
  {\mu} \right)^p$ is equivalent to $V(\phi) = \dfrac {\lambda}
{M_{\text{P}}^{p-4}} \phi^p$ in the literature \cite{chaotic
  Linde}. In this model, the hierarchies (\ref{hierarchies}) do not
hold. We find:
\begin{eqnarray} \label{lf1}
n_S-1 & = & -p(p+2)\left( \dfrac{M_{\rm P}}{\phi}\right)^2\, ;
\nonumber \\
\alpha_S &  = & -2p^2(p+2)\left( \dfrac{M_{\rm P}}{\phi}\right)^4\, ;
\nonumber\\
\beta_S & = & -8p^3(p+2)\left( \dfrac{M_{\rm P}}{\phi}\right)^6\, ;
\nonumber \\
r &  = & 7p^2\left( \dfrac{M_{\rm P}}{\phi}\right)^2\, .
\end{eqnarray}
Note that these quantities are independent of the normalization of the
potential (described by $\Lambda^4/\mu^p$ or, equivalently, by $\lambda$), but
do depend on its shape (described by $p$) as well as on the field value. 

Inflation ends at $\phi_{\rm end}=\dfrac{p M_{\rm P}}{\sqrt{2}}$ where
$\epsilon=1$. Then, it is straightforward to rewrite the inflation
parameters as functions of the number of $e-$folds, $N$:
\begin{eqnarray} \label{lf2}
n_S-1 & = & -\dfrac{2(p+2)}{4N+p}\, ,
\nonumber \\
\alpha_S &  = & -\dfrac{8(p+2)}{(4N+p)^2}=-\dfrac{2}{p+2}(n_S-1)^2\, ,
\nonumber\\
\beta_S & = & -\dfrac{64(p+2)}{(4N+p)^3}=\dfrac{8}{(p+2)^2}(n_S-1)^3\, ,
\nonumber \\
r &  = & \dfrac{14p}{4N+p} = - \dfrac {7p} {p+2} (n_S-1) \, .
\end{eqnarray}

Evidently the spectrum is ``red'' in this model, $n_S-1$, $\alpha_S$ and
$\beta_S$ all being negative. However, this model also cannot accommodate the
current central values, according to which both $|n_S-1|$ and $|\alpha_S|$ are
of order $10^{-2}$. Moreover, $N_{\rm pivot} \leq 50$ implies
$n_S-1 \leq 0.040 \, (0.059)$ for $p = 2\, (4)$, \textit{i.e.} 
$n_S-1$ comes out somewhat below the current central value in this model.

Computing the power directly from Eq.(\ref{power}), we find:
\begin{equation} \label{chaotic_exact}
\mathcal{P}_{\mathcal{R}_c}(N) = \frac {1}{12 \pi^2} \left( \dfrac
{\Lambda^2} {\mu M_{\rm P}} \right)^2 \left[ 2 p \left( N + \dfrac{p}{4}
  \right) \right]^{\dfrac{p+2}{2}} \,.
\end{equation}
This {\em de}creases quickly towards the end of inflation ($N \rightarrow 0$),
again showing that PBH formation is not possible in this model.

\subsubsection{Generalized exponential inflation}

We now turn to the generalized exponential potential \cite{exponential potential}:
\begin{equation} \label{lf21}
V(\phi)=\Lambda^4 e^{\left( \phi/\mu\right) ^p}\, ,
\end{equation}
where $p$ is a positive dimensionless constant and $\mu$ is a constant
with dimension of mass. In this model the hierarchies
(\ref{hierarchies}) among the slow--roll parameters again do not hold
and the values of the spectral parameters depend on the field value
$\phi$:
\begin{eqnarray} \label{lf22}
n_S-1 & = & p\left( \dfrac{M_{\rm P}}{\mu}\right)^2\left[2(p-1)\left(
  \dfrac{\phi}{\mu}\right)^{p-2}-p\left(\dfrac{\phi}{\mu} \right)^{2p-2}
  \right] \, ; 
\nonumber \\
\alpha_S &  = & 2p^2(p-1)\left( \dfrac{M_{\rm P}}{\mu}\right)^4\left[p\left(
  \dfrac{\phi}{\mu}\right)^{3p-4}-(p-2)\left(\dfrac{\phi}{\mu} \right)^{2p-4}
  \right] \, ; 
\nonumber\\
\beta_S & = & 2p^3(p-1)\left( \dfrac{M_{\rm
    P}}{\mu}\right)^6\left[-p(3p-4)\left(
  \dfrac{\phi}{\mu}\right)^{4p-6}+2(p-2)^2\left(\dfrac{\phi}{\mu}
  \right)^{3p-6}   \right] \, ; 
\nonumber \\
r &  = & 7p^2\left( \dfrac{M_{\rm P}}{\mu}\right)^2\left(
\dfrac{\phi}{\mu}\right)^{2p-2}\, .
\end{eqnarray}
We allow $p$ to be a positive real (not necessarily integer) number. If $p$
is not integer, $\phi$ has to be non-negative to get a real potential. In any
case the field $\phi$ will roll from larger to smaller values during
inflation. 

For $p>2$, both terms in the first eq.(\ref{lf22}), or equivalently both
$\epsilon$ and $\eta$, decrease with decreasing $\phi$. The requirements
$|\eta|<1$, $\epsilon<1$ then yield an upper bound on $\phi$, but inflation
will never stop once $\phi$ is below this upper bound. This would require an
additional mechanism to end inflation; we therefore only consider $p<2$ here.

For $p<2$, the requirement $|\eta|<1$ gives a lower bound on $\phi$, which is
approximately given by
\begin{equation}\label{lf23}
\phi_{\rm min}\simeq\mu\left[\dfrac{p |p-1| M_{\rm P}^2}{\mu^2}
  \right]^{\frac{1}{2-p}} \qquad (p<2) \,.
\end{equation}
This bound vanishes for $p=1$. Eqs.(\ref{lf22}) show that this choice leads to
a constant spectral index $n_S$ and vanishing $\alpha_S$ and $\beta_S$. This
means that inflation does not end for $p=1$. Moreover, $\alpha_S=0$ is
(mildly) in conflict with present data, and a constant $n_S$ in the allowed
range will not lead to PBH formation.

If $p>1$, the requirement $\epsilon<1$ implies an upper bound on $\phi$: 
\begin{equation}\label{lf24}
\phi_{\rm max}\simeq\mu\left(\dfrac{\mu}{p M_{\rm P}}
\right)^{\frac{1}{p-1}}\qquad (p>1) \,.
\end{equation}
The number of $e$--folds that occur after the inflaton field had a value
$\phi$ is given by
\begin{equation}\label{lf25}
N(\phi)=\dfrac{1}{p(2-p)}\left( \dfrac{\mu}{M_{\rm
    P}}\right)^2\left[\left(\dfrac{\phi}{\mu} \right)^{2-p}
  -\left(\dfrac{\phi_{\rm min}}{\mu} \right)^{2-p}  \right]\, , 
\end{equation}
where $\phi_{\rm min}$ is given by (\ref{lf23}), and for $p>1$, $\phi$ has to
satisfy $\phi<\phi_{\rm max}$, with $\phi_{\rm max}$ given by
eq.(\ref{lf24}). 

For $p<1$, $n_S$ is always less than $1$, in accord with observation. The
running of the spectral index, given by $\alpha_S$, is also negative, but we
find $\alpha_S>-(n_S-1)^2$ for all allowed combinations of parameters that
allow at least $30$ $e$--folds of inflation after the pivot scale. This model
can therefore not accommodate a sizable and negative value of $\alpha_S$,
either.
\begin{figure}[t]
\vspace {-3cm}
\centering
\includegraphics[width=1.1\textwidth]{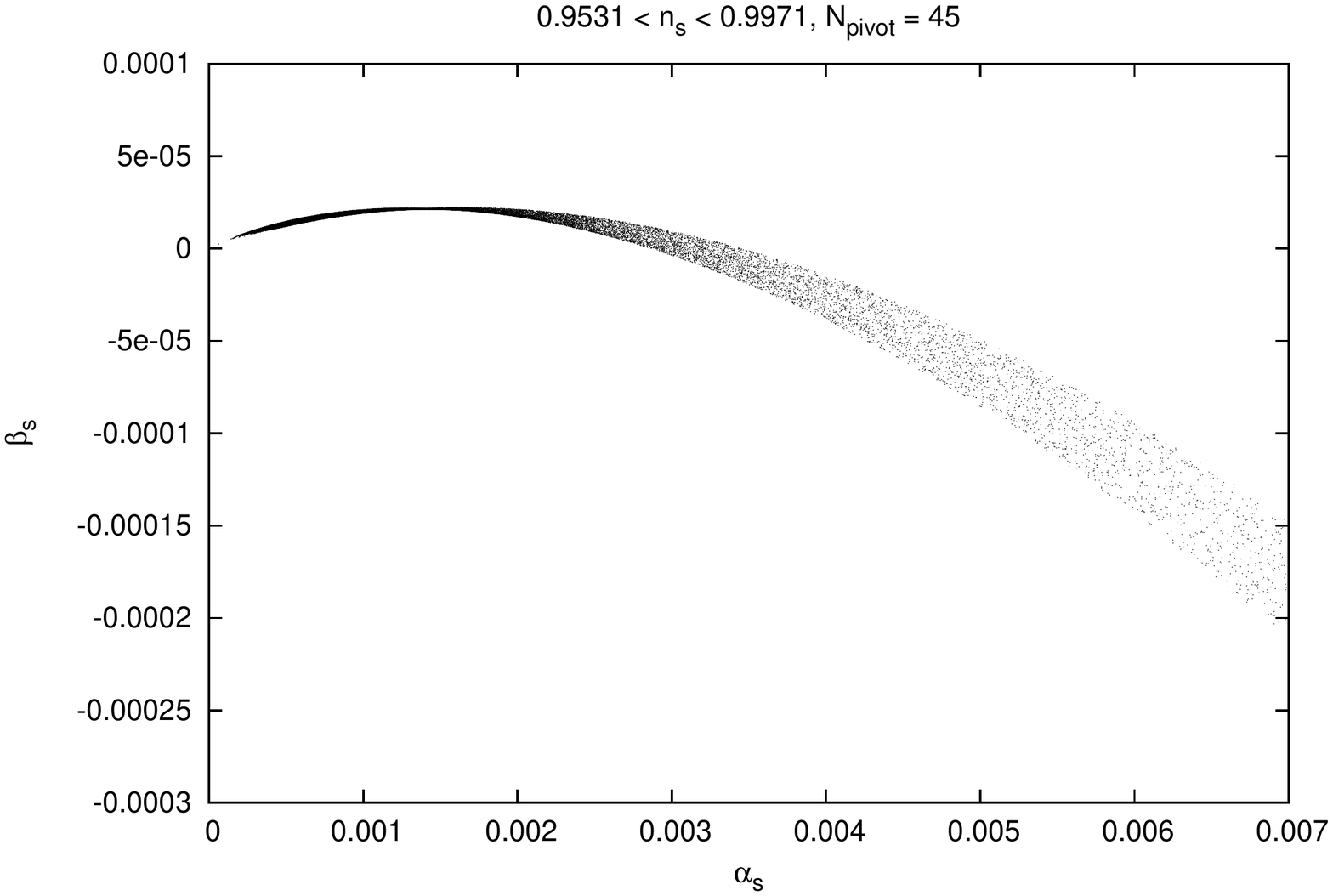}
\vspace {-1.5cm}
\caption{Scatter plot of allowed values of $\alpha_S$ and $\beta_S$ assuming
  that $n_S$ lies in its currently allowed $2\,\sigma$ range and $45$
  $e$--folds of inflation occurred after the pivot scale for potential
  (\ref{lf21}).} 
\label{fig:scatter}
\end{figure}

For $1<p<2$, $n_S-1$ can have either sign, while $\alpha_S$ is always
positive, in contrast to the bound of current data. Quite large and positive
$\alpha_S$ are in principle possible, if parameters are chosen such that the
two contributions to $n_S-1$ in eq.(\ref{lf22}) cancel approximately. However,
such large values of $\alpha_S$ are definitely in conflict with
observation. Moreover, if $\alpha_S>0.004$, $\beta_S$ turns negative, limiting
the growth of power at small scales. This is illustrated in
figure~\ref{fig:scatter}, which shows a scatter plot of allowed values of
$\alpha_S$ and $\beta_S$ assuming that $n_S$ lies in its currently allowed
$2\,\sigma$ range and $45$ $e$--folds of inflation occurred after the pivot
scale.

Using the exact expression (\ref{power}) we numerically find $n \leq 1.15$ at
scales relevant for the formation of $10^{15}$ g PBHs; we saw in section 2
that $n>1.37$ is required for the formation of such PBHs. This model therefore
cannot accommodate PBH formation, either.

\subsubsection{Inflation with negative exponential and Higgs inflation}

Another potential that has been proposed is \cite{review}
\begin{equation}\label{lf3}
V(\phi) = V_0\left( 1-e^{-q\phi/M_{\text{P}}} \right )\, .
\end{equation}

For $q>0$ the inflaton field $\phi$ rolls towards smaller field values
during inflation.\footnote{If negative values of $\phi$ are allowed,
  the potential (\ref{lf3}) becomes unbounded for $\phi \rightarrow
  -\infty$. In this case additional terms have to be added to the
  potential, which we again assume to be unimportant during the
  slow--roll phase.} The potential is sufficiently flat only for
$q\phi>M_{\rm P}$, where the hierarchies (\ref{hierarchies}) between
slow--roll parameters hold. The inflationary parameters are
\begin{eqnarray} \label{lf32}
n_S - 1 & \simeq & -2 q^2 e^{-q\phi/M_{\text{P}}} \, ;
\nonumber \\
\alpha_S & \simeq & -2 q^4 e^{-2q\phi/M_{\text{P}}}=-\dfrac{1}{2}(n_S-1)^2 \, ;
\nonumber \\
\beta_S & \simeq & -4 q^6 e^{-3q\phi/M_{\text{P}}}=\dfrac{1}{2}(n_S-1)^3\, ;
\nonumber \\
r & \simeq & 7 q^2 e^{-2q\phi/M_{\text{P}}}\, ,
\end{eqnarray}
where we have approximated the denominators of eqs.(\ref{slow-roll})
by $V_0$; this is appropriate for the phase of slow--roll where $n_S
\simeq 1$, unless $q^2 \ll 1$. Inflation ends at $\phi_{\text{end}} =
2M_{\rm P} \dfrac{ {\rm ln}\,q} {q} $. $N$ $e-$folds before the end of
inflation the spectral parameters are given by
\begin{eqnarray} \label{lf33}
n_S - 1 & \simeq & -\dfrac{2}{N+1}  \, ,
\nonumber \\
\alpha_S & \simeq & -\dfrac{2}{(N+1)^2} \, ,\\
\nonumber \\
\beta_S & \simeq & -\dfrac{4}{(N+1)^3}\, ,\\
\nonumber \\
r & \simeq & \dfrac{7}{q^2}\dfrac{1}{(N+1)^2}\, .
\end{eqnarray}
Note that the $q$--dependence cancels when the spectral parameters are
expressed in terms of $N$. 

In this model $n_S-1$ and $\alpha_S$ are manifestly negative, in agreement
with current data. However, while $n_S-1$ also has approximately the right
magnitude, $|\alpha_S|$ at the pivot scale is much smaller than the
experimental central value. Moreover, since $\alpha_S$ and $\beta_S$ are both
negative, PBH formation is not possible in this model; this can also be seen
from the exact expression (\ref{power}), which shows that the power always
decreases with decreasing $N$.

A very similar potential describes the Higgs inflation model
\cite{Higgs_inflation} where the Higgs boson of the Standard Model
(SM) plays the role of the inflaton. Starting point of this model is
the non--minimal coupling of the Higgs field to gravity. The relevant
part of the action in the Jordan frame is:
\begin{equation} \label{Higgs1}
S_{\rm J} = \int {\rm d}^4 x\sqrt{-g}\left\lbrace -\dfrac{M^2+\xi h^2}{2}R +
\dfrac{\partial_\mu h\partial^\mu h} {2} - \dfrac {\lambda} {4} (h^2-v^2)^2
\right\rbrace\, ,   
\end{equation}
where $M$ is some mass parameter\footnote{In the range of $\xi$ of
  interest to us, $M \simeq M_{\rm P}$ \cite{Higgs_inflation}.}, $R$ is
the scalar curvature, $h$ is the Higgs field in the unitary gauge and
$\xi$ determines the coupling of the Higgs to gravity.\footnote{Higgs
  inflation requires $\xi \gg 1$. This leads to a breakdown of
  tree--level unitarity at scales well below the Planck scale
  \cite{higgs1}, but according to ref.\cite{higgs2} this does not
  invalidate the scenario, since the relevant energy scale during
  inflation always remains in the unitary regime.} By making a
conformal transformation from the Jordan frame to the Einstein frame
one can get rid of the non--minimal coupling:
\begin{equation} \label{Higgs2}
\hat{g}_{\mu\nu}=\Omega^{2}g_{\mu\nu}\, , \qquad \Omega^{2}=1+\frac{\xi
  h^{2}}{M^{2}_{\rm P}}\, . 
\end{equation}
This transformation induces a non--canonical kinetic energy term for $h$. 
It is therefore convenient to redefine $h$ in terms of the scalar field $\phi$
which casts the kinetic term into the canonical form \cite{Higgs_inflation}:
\begin{equation} \label{Higgs3}
\frac{d\phi}{dh} = \sqrt{ \frac {\Omega^{2} + 6 \xi^{2} h^{2} / M^{2}_{\rm P}}
  {\Omega^{4}} }\, . 
\end{equation}
In terms of this new field, the potential is
\begin{equation} \label{Higgs4}
V(\phi) = \frac {1} {\Omega(\phi)^{4}} \frac {\lambda} {4} \left[ h(\phi)^{2}
- v^{2} \right] ^{2}\, . 
\end{equation}
For small field value, $h^2 \ll M_{\rm P}^2/\xi$, one has $h \simeq
\phi$ and $\Omega^2 \simeq 1$; the two frames are indistinguishable so
the potential for the field $\phi$ is the same as that for the initial
Higgs field. However, for large values, $h \gg M_{\rm P}/\sqrt{\xi}$,
one has $\Omega^2 \simeq \xi h^2 / M_{\rm P}^2$, and
\cite{Higgs_inflation}
\begin{equation} \label{Higgs5}
h \simeq \frac {M_{\rm P}} {\sqrt{\xi}} \exp \left( \frac {\phi}
  {\sqrt{6} M_{\rm P} } \right)\, . 
\end{equation}
Substituting this into eq.(\ref{Higgs4}) we obtain the expression for the
potential
\begin{equation} \label{Higgs6}
V(\phi) = \frac{ \lambda M_{\rm P}^{4}} {4{\xi}^{2}} \left[ 1 -\exp\left(
-\frac {2\phi} {\sqrt{6} M_{\rm P}} \right)\right]^{2}\, . 
\end{equation}

Recall that this expression holds only for $h \gg M_{\rm P}/\sqrt{\xi}$, which
implies that the exponential term in eq.(\ref{Higgs6}) is small. The square of
this term is then even smaller, and can be neglected during inflation. The
potential therefore effectively almost reduces to the form (\ref{lf3}), with
$q = 2/\sqrt{6}$, except that the exponential term is multiplied with 2 (due
to the square in eq.(\ref{Higgs6})). This also increases $n_S-1$ by a factor of  2:
\begin{equation} \label{Higgs7}
n_S-1 \simeq  -\dfrac{8}{3}\dfrac{M_{\rm P}^2}{\xi h^2}\, .
\end{equation}
However, this factor of 2 cancels if $n_S-1$ is expressed in terms of the
number of $e-$folds of inflation that occur after the field had the value
$\phi$, \textit{i.e.} the first eq.(\ref{lf33}) remains valid in Higgs
inflation. Moreover, $\alpha_S$ and $\beta_S$ are as in eqs.(\ref{lf32}) when
expressed in terms of $n_S - 1$ or $N$. For most practical purposes (of
inflation), Higgs inflation can therefore be understood as a particle physics
implementation of negative exponential inflation.

\subsubsection{Natural inflation}

One way to obtain a very flat potential is to consider the natural inflation
\cite{natural inflation} where a Pseudo Nambu--Goldstone Boson (PNGB) is used
as inflaton.\footnote{Natural inflation can be either a large-- or
  small--field inflation model, depending on the value of $f$. Here we assume
  $f> M_{\rm P}$.} In this model, the inflaton field has a particular form of
the potential which results from explicit breaking of a shift symmetry:
\begin{equation}\label{natural}
V(\phi)=\Lambda^4\left[1\pm\text{cos}\left(a\dfrac{\phi}{f} \right)  \right]\,
, 
\end{equation}
We will take the positive sign in eq.(\ref{natural}) and assume that initially
$\phi \ll f/a$. For appropriately chosen values of the mass scales, \textit{e.g.} $f/a
\sim M_{\text{P}}$ and $\Lambda \sim M_{\text{GUT}}\sim 10^{16}$ GeV, the PNGB
field $\phi$ can drive inflation. We set $a=1$ for simplicity and treat $f$ as
a free parameter. The Slow--roll parameters are then given by
\begin{eqnarray} \label{natural1}
\epsilon & = & \dfrac{1}{2}\left(
\dfrac {M_{\text{P}}} {f} \right)^2 \left[ \dfrac {\sin(\phi/f)} {1 +
    \cos(\phi/f)} \right]^2 \simeq \dfrac{1}{8} \left( \dfrac{M_{\text{P}}}
       {f} \right)^2 \left( \dfrac{\phi}{f} \right)^2\,  ; 
 \nonumber \\
\eta & = & -\left( \dfrac {M_{\text{P}}} {f} \right)^2 \left[\dfrac{
    \cos(\phi/f)} { 1 + \cos(\phi/f) } \right] \simeq -\dfrac {1} {2} \left(
\dfrac{M_{\text{P}}} {f} \right)^2\, ; 
 \nonumber\\
\xi^2 & = & -\left( \dfrac {M_{\text{P}}}{f} \right)^4 \left[ \dfrac
  {\sin(\phi/f)} { 1 + \cos(\phi/f) } \right]^2 \simeq -\dfrac {1} {4} \left(
\dfrac {M_{\text{P}}} {f} \right)^4 \left( \dfrac{\phi}{f} \right)^2\, ; 
\nonumber\\
\sigma^3 & = & \left( \dfrac{M_{\text{P}}} {f} \right)^6 \dfrac {
  \cos(\phi/f)\, \sin^2(\phi/f) } { \left[ 1 + \text{cos}(\phi/f) \right]^3}
\simeq \dfrac{1}{8} \left( \dfrac{M_{\text{P}}}{f} \right)^6 \left(
\dfrac{\phi}{f}\right)^2\,  ,
\end{eqnarray}
where the approximate equalities hold for $\phi \ll f$.  It is clear
that the hierarchies (\ref{hierarchies}) do not hold among the
slow--roll parameters\footnote{The first strong inequality in
  (\ref{hierarchies}) does hold for $\phi \ll f$, but the second one
  does not even hold in this limit.}, and we find the following
inflation parameters:
\begin{eqnarray} \label{natural2}
n_S-1 & = & -\left( \dfrac {M_{\rm P}}{f} \right)^2 \dfrac{ 3 - \cos(\phi/f) }
{ 1 + \cos(\phi/f) }\, ; 
 \nonumber \\
\alpha_S & = & -4 \left( \dfrac{M_{\rm P}} {f} \right)^4 \dfrac { 1 -
  \cos(\phi/f) } { \left[ 1 + \cos(\phi/f)\right] ^2}\, ;
 \nonumber\\
\beta_S & = & -4 \left( \dfrac {M_{\rm P}} {f} \right)^6 \dfrac{ \left[ 1 -
    \cos(\phi/f) \right] \left[ 3 - \cos(\phi/f) \right] } { \left[ 1 +
    \cos(\phi/f) \right] ^3}\, ;
\nonumber\\
r & = & 7 \left( \dfrac {M_{\rm P}} {f}\right)^2 \dfrac{ 1 - \cos(\phi/f) } {
  1 + \cos(\phi/f)}\, .
\end{eqnarray}
Inflation ends at $|\eta|=1$ and the relation between the inflaton
field and the number of $e$--folds is given by
\begin{equation} \label{natural3}
\cos\left( \dfrac{\phi}{f}\right) = 1 - y\, ,
\end{equation}
where $y\equiv\dfrac{x^2+2}{x^2+1}e^{-Nx^{2}}$, $x\equiv\dfrac{M_{\rm
    P}}{f}$. Inserting eq.(\ref{natural3}) into (\ref{natural2}) yields
\begin{eqnarray} \label{natural4}
n_S-1 & = & -x^{2}\, \dfrac{2+y}{2-y}\, ,
\nonumber \\
\alpha_S & = & -4x^{4}\, \dfrac{y}{(2-y)^2}\, ,
\nonumber\\
\beta_S & = & -4x^{6}\, \dfrac{y(2+y)}{(2-y)^3}\, ,
\nonumber\\
r & = &  7x^{2}\, \dfrac{y}{2-y}\, .
\end{eqnarray}

In the course of inflation $y$ increases from a rather small value to $y_{\rm
  end}=\dfrac{2+x^2}{1+x^2}$ at the end of inflation
$(N=0)$. Eqs.(\ref{natural4}) show that $n_S-1$, $\alpha_S$ and $\beta_S$
become more negative as $y$ increases, indicating that the power is reduced at
smaller scales. This can also be seen from the exact expression (\ref{power})
which gives
\begin{equation}
\mathcal{P}_{\mathcal{R}_c}=\dfrac{\Lambda^4f^2}{12\pi^2M_{\rm
    P}^6}\dfrac{(2-y)^2}{y}\, , 
\end{equation}
which decreases with increasing $y \in [0,2]$. PBH formation is therefore not
possible in this model. 

For fixed $N$, the spectral parameters are determined by $x$. $|n_S-1|$ can
clearly be made as large as desired (with $n_S < 1$) by choosing a large value
of $x$, \textit{i.e.} a small value of $f$. On the other hand, $|\alpha_S|$
reaches a maximum at $x^2\simeq2/N$ for $y\ll1$; note that $y$ decreases with
increasing $x^2$. This gives $\alpha_S\gtrsim-1.5/N^2$, \textit{i.e.}
$|\alpha_S|$ at the pivot scale cannot be larger than $10^{-3}$ in this model,
well below the current central value.

\subsubsection{Arctan inflation}

Another inflation model which we are interested to study has been introduced
in \cite{arctan}:
\begin{equation} \label{arc1}
V(\phi)=V_0\left[1+\dfrac{2}{\pi}\rm {arctan}\left(\dfrac{\phi}{\mu} \right)
  \right]\, .
\end{equation}
This model allows inflation with $n_S\simeq1$ if $\mu\gg M_{\rm P}$ or
$\phi\gg \mu$. However, inflation can be ended by the potential
(\ref{arc1}) only if $\mu\lesssim0.8 M_{\rm P}$, since otherwise
$\epsilon,|\eta|<1\quad \forall\, \phi$. A finite period of inflation
thus requires that $\phi\gg\mu$ initially; at the end of inflation,
$\phi\rightarrow-\infty$, \textit{i.e.}  $V\rightarrow0$. During the
slow--roll phase the hierarchies (\ref{hierarchies}) between the
slow--roll parameters hold, and we find
\begin{eqnarray} \label{arc2}
n_S-1 & \simeq & -\dfrac{4}{\pi}\left(\dfrac{M_{\rm P}}{\phi} \right)^2
\dfrac{\mu}{\phi}\, , 
\nonumber \\
\alpha_S & \simeq & -\dfrac{12}{\pi^2}\left(\dfrac{M_{\rm P}}{\phi} \right)^4
\left( \dfrac{\mu}{\phi}\right)^2 = -\dfrac{3}{4}(n_S-1)^2\, , 
\nonumber \\
\beta_S & \simeq & -\dfrac{72}{\pi^3}\left(\dfrac{M_{\rm P}}{\phi} \right)^6
\left( \dfrac{\mu}{\phi}\right)^3 = \dfrac{9}{8}(n_S-1)^3\, , 
\nonumber \\
r & \simeq & \dfrac{7}{\pi^2}\left(\dfrac{M_{\rm P}}{\phi} \right)^2
\left(\dfrac{\mu}{\phi} \right)^2\, , 
\end{eqnarray}
where we have approximated the denominators of eqs.(\ref{slow-roll}) by $2
V_0$, as appropriate for the slow--roll phase where $\phi \gg \mu$.
In terms of the number $N$ of $e$--folds of inflation that occurred after the
inflaton field reached the value $\phi$, we find
\begin{equation}\label{arc3}
n_S - 1 \simeq - \dfrac {4} {3N+\pi} \, ,
\end{equation}
where we have used the fact that inflation ends at $\phi_{\rm
  end}\simeq\left(M_{\rm P}^2\mu \right)^{1/3}$. This agrees with the
currently allowed range for $38\leq N \leq 95$. However, while $\alpha_S$ is
negative, its absolute value is only of order $10^{-3}$ for allowed values of
$n_S$; moreover, since $\beta_S$ is also negative, PBH formation is not
possible. Indeed, one can see from the exact expression (\ref{power}) that the
power decreases steadily during inflation:
\begin{equation} \label{arc4}
\mathcal{P}_{\mathcal{R}_c} = \dfrac {V_0} {12 \pi^2 M_{\rm P}^6} \left[ 1 +
  \dfrac {2} {\phi} \arctan \left( \dfrac {\phi} {\mu} \right) \right]^3
\left( 1 + \dfrac {\phi^2} {\mu^2} \right)\,.
\end{equation}
This decreases with decreasing $\phi$ for $\phi_{\rm end} \leq \phi < \infty$.

\section{Summary and Conclusions}

In this paper we reviewed the formation of primordial black holes
using the Press--Schechter formalism. We found that the formation of
PBHs with mass larger than $10^{15}$ g, whose lifetime exceeds the age
of the Universe, will be produced at sufficient abundance to form the
cold Dark Matter if the spectral index at scale $k_{\rm PBH}$ is about
$1.37$ for the threshold value $\delta_{\text{th}}=1/3$. This spectral
index is much above the value measured at much larger length scales in
the CMB. PBH formation therefore requires significant positive running
of the spectral index when $k$ is increased. 

We compared this with the values of the spectral index and its running
derived from current data on large scale structure. These include
analyses of CMB anisotropies from the WMAP (7 year) and SPT
collaborations, as well as data on baryonic acoustic oscillations and
on the abundance of clusters, and direct measurements of the Hubble
constant $H_0$. At the pivot scale of this data set one finds
$n_S(k_{\rm pivot}) = 0.9751$ as central value. The first derivative
$\alpha_S(k_0)$ would then need to exceed $0.020$ if it alone were
responsible for the required increase of the spectral index; this is
more than $3\,\sigma$ above the current central value of this
quantity. However, the second derivative (the ``running of the
running'') of the spectral index is currently only very weakly
constrained. We showed in a model--independent analysis that this
easily allows values of $n(k_{\rm PBH})$ large enough for PBH
formation, even if the first derivative of the spectral index is
negative at CMB scales.

In section~3 we applied this formalism to a wide class of inflationary
models, under the constraints imposed by the data mentioned above. We
classified the inflation models in small--field and large--field
models. We have shown that only one small--field model, the
running--mass model, allows sizable positive running of running of the
spectral index, and is thus a good candidate for long--lived PBHs
formation, albeit only in a narrow range of parameter space. In
contrast, all the large--field models we studied predict small or
negative values for the second derivative of the spectral index, and
thus predict negligible PBH formation due to the collapse of overdense
regions seeded during inflation. Recall, however, that other
mechanisms of PBH formation have been suggested, \textit{e.g.} in the
``waterfall'' phase of models of hybrid inflation, or during first
order phase transitions.

As a by--product of our analysis, we found that most of the models we
studied either predict $n_S < 1$, as indicated by present data, or can
at least accommodate it, the single exception being inverse power law
inflation (a large field model). In contrast, {\em none} of the models
we analyzed allows to reproduce a large negative value of $\alpha_S$,
as preferred by current data; this confirms the results of the general
analysis of ref.\cite{Peiris}. If future data confirm with high
precision that $\alpha_S \lsim -0.01$, all simple single--field models
of inflation would be excluded. Similarly, proving conclusively that
the second derivative of the spectral index is positive would exclude
all the large--field models we investigated. Future analyses of the
spectrum of primordial density perturbations thus hold great promise
to discriminate between inflationary scenarios, or even to challenge
the paradigm of single--field inflation.

\section*{Acknowledgments}

This work was supported by the TR33 ``The Dark Universe'' funded by
the Deutsche Forschungsgemeinschaft. EE also thanks the Bonn--Cologne
Graduate School for support.


\begin{thebibliography}{99}

\bibitem{kt}
For an introduction into the problems of standard cosmology and their
solution by inflation, as well as a review of the early literature,
see E.~W.~Kolb and M.~S.~Turner, \textit{The Early Universe},
Addison--Wesley (1990).

\bibitem{Lyth Book} 
D.~H.~Lyth and A.~R.~Liddle, \textit{The Primordial Density Perturbation},
Cambridge University Press (2009).

\bibitem{review}
For reviews on inflationary models, see
D.~Lyth and A.~Riotto, \textit{Phys.~Rept.} {\bf 314} (1999) 1-146 [arXiv:
  hep-ph/9807278]; A.~Mazumdar and J.~Rocher, \textit{Phys.~Rept.}
\textbf{497} (2011) 85 [arXiv: 1001.0993 [hep-ph]].

\bibitem{Carr1}
B.~J.~Carr and S.~W.~Hawking, \textit{Mon.~Not.~Roy.~Astron.~Soc.}
\textbf{168} (1974) 399.

\bibitem{Carr2}
B.~J.~Carr, \textit{Astrophys.~J.} \textbf{201} (1975) 1.

\bibitem{PBH_formation}
B.~J.~Carr and J.~E.~Lidsey, \textit{Phys.~Rev.} \textbf{D48} (1993) 543;
B.~J.~Carr, J.~H.~Gilbert and J.~E.~Lidsey \textit{Phys.~Rev.} \textbf{D50}
(1994) 4853 [arXiv: astro-ph/9405027].

\bibitem{encieh}
M.~Drees and E.~Erfani, \textit{JCAP} \textbf{1104} (2011) 005 [arXiv:
  1102.2340 [hep-ph]].

\bibitem{pbh_constraint}
A.~M.~Green and A.~R.~Liddle, {\it Phys.~Rev.} {\bf D56} (1997) 6166 [arXiv: astro-ph/9704251];
H.~I.~Kim, C.~H.~Lee and J.~H.~MacGibbon, {\it Phys.~Rev.} {\bf D59}
(1999) 063004 [arXiv: astro-ph/9901030];
P.~Pina, {\it Phys.~Rev.} {\bf D72} (2005) 124004 [arXiv: astro-ph/0510052];
E.~Bugaev and P.~Klimai, {\it Phys.~Rev.} {\bf D79} (2009) 103511
[arXiv: 0812.4247 [astro-ph]];
A.~S.~Josan, A.~M.~Green and K.~A.~Malik, {\it Phys.~Rev.} {\bf D79} (2009) 103520
[arXiv: 0903.3184 [astro-ph.CO]];
K.~Kohri, D.~H.~Lyth, A.~Melchiorri, \textit{JCAP} \textbf{0804} (2008) 038 [arXiv: 0711.5006 [hep-ph]];
L.~Alabidi and K.~Kohri, \textit{Phys.~Rev.} \textbf{D80} (2009) 063511 [arXiv: 0906.1398 [astro-ph.CO]].

\bibitem{earlyPBH}
D.~H.~Lyth, K.~A.~Malik, M.~Sasaki and I.~Zaballa, \textit{JCAP} \textbf{0601}
(2006) 011 [arXiv: astro-ph/0510647]; 
I.~Zaballa, A.~M.~Green, K.~A.~Malik and M.~Sasaki, \textit{JCAP}
\textbf{0703} (2007) 010 [arXiv: astro-ph/0612379].

\bibitem{WMAP7} 
E.~Komatsu \textit{et. al.}, \textit{Astrophys.~J.~Suppl.} \textbf{192} (2011)
18 [arXiv: 1001.4538 [astro-ph.CO].

\bibitem{SPT}
R.~Keisler \textit{et. al.}, arXiv: 1105.3182 [astro-ph.CO].

\bibitem{BAO}
W.~J.~Percival {\it et al.}, {\it MNRAS} {\bf 401} (2010) 2148.

\bibitem{H0}
A.~G.~Riess {\it et al.}, {\it ApJ} {\bf 730} (2011) 119.

\bibitem{CLUSTERS}
A.~Vikhlinin \textit{et. al.}, \textit{ApJ} \textbf{692} (2009) 1060.

\bibitem{Press-Schechter} 
W.~H.~Press and P.~Schechter, \textit{Astrophys.~J.} \textbf{187} (1974) 425.

\bibitem{Zeldovich}
Y.~B~. Zel'dovich and I.~.D~. Novikov, \textit{azh} \textbf{43} (1966) 758;
Y.~B.~Zel'dovich and I.~D.~Novikov, \textit{Sov.~Astron.~A.~J.} \textbf{10}
(1967) 602. 

\bibitem{Hawking1}
S.~W.~Hawking, \textit{Mon.~Not.~Roy.~Astron.~Soc.} \textbf{152} (1971) 75.

\bibitem{khlopov}
M.~Y.~Khlopov and A.~G.~Polnarev, \textit{Phys.~Lett.} \textbf{B97} (1980)
383. 

\bibitem{Hawking2}
S.~W.~Hawking, \textit{Phys.~Lett.} \textbf{B231} (1989) 237.

\bibitem{bubble}
M.~Crawford and D.~N.~Schramm, \textit{Nature} \textbf{298} (1982) 538;
I.~G.~Moss, \textit{Phys.~Rev.} \textbf{D50} (1994) 676. 

\bibitem{domain walls}
S.~G.~Rubin, M.~Y.~Khlopov and A.~S.~Sakharov, \textit{Grav.~Cosmol.}
\textbf{6} (2001) 1; V.~I.~Dokuchaev, Y.~N.~Eroshenko and S.~G.~Rubin,
\textit{Grav.~Cosmol.} \textbf{11} (2005) 99 [arXiv: astro-ph/0412418]. 

\bibitem{Jedamzik}
J.~Niemeyer and K.~Jedamzik, \textit{Phys.~Rev.~Lett.} \textbf{80} (1998)
5481; \textit{Phys.~Rev.} \textbf{D59} (1999) 124013. 

\bibitem{Josan}
A.~S.~Josan and A.~M.~Green, \textit{Phys.~Rev.} \textbf{D82} (2010) 047303
[arXiv: 1004.5347 [hep-ph]].

\bibitem{Kosowsky}
A.~Kosowsky and M.~S.~Turner, \textit{Phys.~Rev.} \textbf{D52} (1995) 1739;
N.~D\"{u}chting, \textit{Phys.~Rev.} {\bf D70} (2004) 064015 [arXiv:
astro-ph/0406260]; Q.-G.~Huang, \textit{JCAP} \textbf{0611} (2006) 004
[arXiv: astro-ph/0610389].

\bibitem{Hawking3} 
S.~W.~Hawking, \textit{Mon.~Not.~R.~Astron.~Soc.} \textbf{168} (1974) 399.

\bibitem{observation}
B.~Carr, K.~Kohri, Y.~Sendouda and J.~Yokoyama, \textit{Phys.~Rev.}
\textbf{D81} (2010) 104019 [arXiv: astro-ph/0912.5297].

\bibitem{correlation}
M.~Cort\^es, A.~R.~Liddle and P.~Mukherjee, \textit{Phys.~Rev.} \textbf{D75}
(2007) 083520 [arXiv: astro-ph/0702170].

\bibitem{running of running}
Q.~Huang, \textit{JCAP} \textbf{0611} (2006) 004 [arXiv: astro-ph/0610389].

\bibitem{tensor}
S.~Hannestad, S.~H.~Hansen, F.~L.~Villante, \textit{Astropart.~Phys.}
\textbf{16} (2001) 137 [arXiv: astro-ph/0012009].

\bibitem{running}
H.~V.~Peiris and R.~Easther, \textit{JCAP} \textbf{0807} (2008) 024 [arXiv: 0805.2154 [astro-ph]].

\bibitem{e-folds}
A.~R.~Liddle and S.~M.~Leach, \textit{Phys.~Rev.} \textbf{D68} (2003) 103503
[arXiv: astro-ph/0305263].

\bibitem{Hybrid Linde}
A.~D.~Linde, \textit{Phys.~Rev.} \textbf{D49} (1994) 748 [arXiv:
  astro-ph/9307002].

\bibitem{PBHs hybrid}
D.~H.~Lyth, \textit{JCAP} \textbf{1107} (2011) 035 [arXiv: 1012.4617 [astro-ph.CO]]; J.~Garcia-Bellido, A.~Linde and
D.~Wands, \textit{Phys.~Rev.} \textbf{D54} (1996) 6040 [arXiv:
  astro-ph/9605094v3]; T.~Kanazawa, M.~Kawasaki and T.~Yanagida,
\textit{Phys.~Lett.} \textbf{B482} (2000) 174 [arXiv: hep-ph/0002236v2].

\bibitem{Hilltop inflation}
P.~Adshead, R.~Easther, J.~Pritchard and A.~Loeb, \textit{JCAP} \textbf{1102} (2011) 021 [arXiv: 1007.3748v3 [astro-ph.CO]].

\bibitem{Stewart}
E.~D.~Stewart, \textit{Phys.~Lett.} \textbf{B391} (1997) 34  [arXiv:
  hep-ph/9606241]; E.~D.~Stewart, \textit{Phys.~Rev.} \textbf{D56} (1997) 2019
[arXiv: hep-ph/9703232].

\bibitem{dynamical SUSY}
G.~F.~Giudice, R.~Rattazzi, \textit{Phys.~Rept.} \textbf{322} (1999) 419
[arXiv: hep-ph/9801271].

\bibitem{gravitational waves}
D.~H.~Lyth, \textit{Phys.~Rev.~Lett.} \textbf{78} (1997) 1861 [arXiv:
  hep-ph/9606387].

\bibitem{chaotic Linde}
A.~D.~Linde, \textit{Phys.~Lett.} \textbf{B129} (1983) 177.

\bibitem{exponential potential}
A.~Kosowsky and M.~S.~Turner, \textit{Phys.~Rev.} \textbf{D52} (1995) 1739-1743 [arXiv: astro-ph/9504071v2].

\bibitem{Higgs_inflation}
F.~L.~Bezrukov and M.~Shaposhnikov, \textit{Phys.~Lett.} \textbf{B659} (2008)
703 [arXiv: 0710.3755 [hep-th]].

\bibitem{higgs1}
C.~P.~Burgess, H.~M.~Lee and M.~Trott, \textit{JHEP} {\bf 0909} (2009) 103
[arXiv: 0902.4465 [hep-ph]] and \textit{JHEP} {\bf 1007} (2010) 007 [arXiv: 1002.2730
[hep-ph]].

\bibitem{higgs2}
F.~Bezrukov, A.~Magnin, M.~Shaposhnikov, S.~Sibiryakov, \textit{JHEP} {\bf
  1101} (2011) 016 [arXiv: 1008.5157 [hep-ph]].

\bibitem{natural inflation} 
K.~Freese, J.~A.~Frieman and A.~V.~Olinto, \textit{Phys.~Rev.~Lett.}
\textbf{65} (1990) 3233.

\bibitem{arctan}
S.~M.~Leach, A.~R.~Liddle, J.~Martin, D.~J.~Schwarz, \textit{Phys.~Rev.}
\textbf{D66} (2002) 023515 [arXiv: astro-ph/0202094].

\bibitem{Peiris}
R.~Easther and H.~Peiris, \textit{JCAP} {\bf 0609} (2006) 010 [arXiv:
  astro-ph/0604214].

\end{thebibliography}
\end{document}